# QAC: Quantum-computing Aided Composition

Omar Costa Hamido, PhD

ICCMR, University of Plymouth, UK

## Abstract

In this chapter I will discuss the role of quantum computing in computer music and how it can be integrated to better serve the creative artists. I will start by considering different approaches in current computer music and quantum computing tools, as well as reviewing some previous attempts to integrate them. Then, I will reflect on the meaning of this integration and present what I coined as QAC (Quantum-computing Aided Composition) as well as an early attempt at realizing it. This chapter will also introduce *The QAC Toolkit* Max package,[1] analyze its performance, and explore some examples of what it can offer to realtime creative practice. Lastly, I will present a real case scenario of QAC in the creative work *Disklavier Prelude #3*.

## Computer Music and Quantum Computing tools

Recent literature exploring the intersection of Quantum Computing (QC) with creative music practice, of which this book is a prime example, have welcomed the potential increase in speed and computational power offered by future fault tolerant QC machines. In the context of Computer Music (CM), a need for more powerful machines is evident in the limitations that classical machines still present for the realtime (or near realtime) control of compositional processes [2].

Several research projects have already proposed proof-of-concept implementations that integrate QC with music practice, in simulation or with current hardware. However, there is still no consistency between the tools, and approaches undertaken in these explorations, and current CM practice. More importantly, a disconnect between scientific research and creative practice, may only serve to maintain a gap between scientists and artists. My proposed line of inquiry here intends to bridge that gap by focusing on the tools used and how they articulate with realtime creative music practices.

[1] *The QAC Toolkit* is available via the Max Package Manager [1].

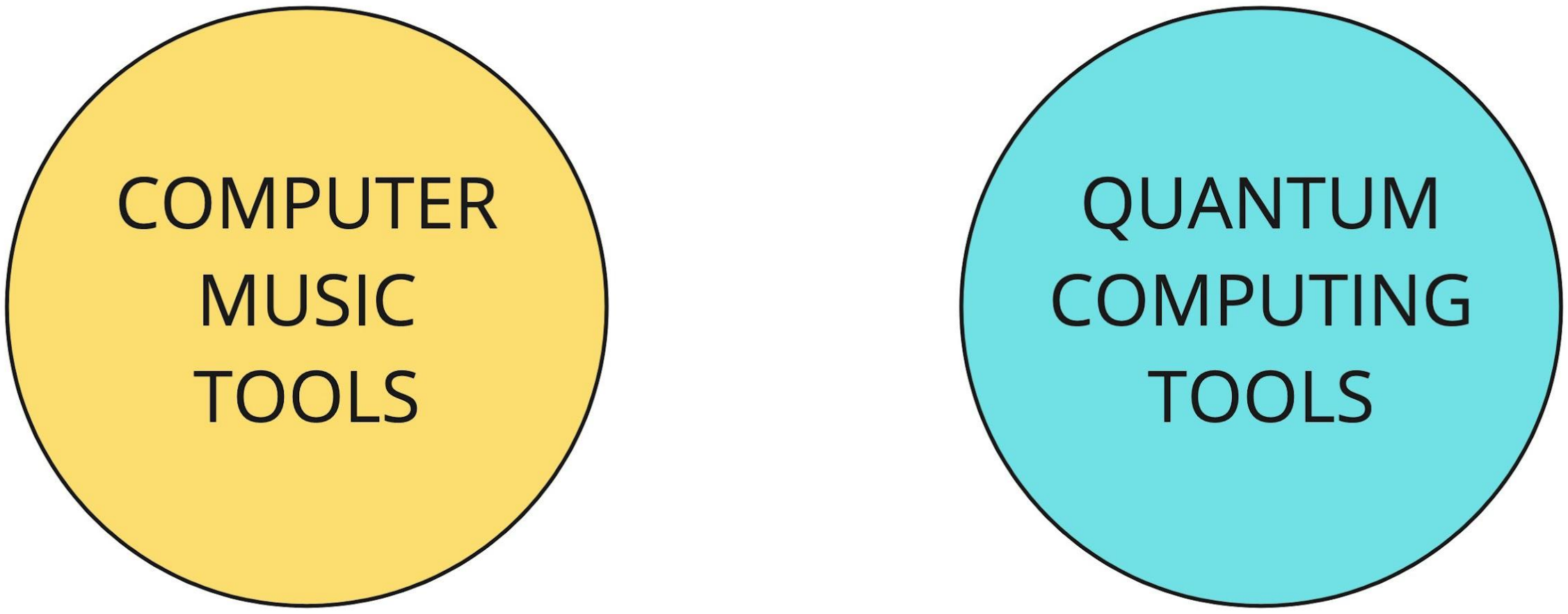

Figure 1 - Computer Music Tools and Quantum Computing Tools

Modern Computer Music tools and Quantum Computing tools have been shaped by their main users in their practice in such a way that each might currently be regarded as capable of drawing their own self-contained world of practice and practitioners (see figure 1). CM tools include score engravers, Digital Audio Workstations (DAW), and visual programming environments, like Musescore [3], Ableton Live [4], and Max/MSP[2] [5], respectively. The use of these 3 categories of tools is deeply embedded in the creative practice of writing musical scores, recording and producing a track, and developing new interactive instruments as well as enabling realtime control of compositional processes.

On the other hand, current QC tools seem to inherit greatly from code-based programming practices, where the majority of its user base is found. These include the different QC programming frameworks[3] like Qiskit [6], Cirq [7], and pyQuil [8], that are accessed using a terminal or a Jupyter notebook, as well as some web apps that allow the design of circuits online like Strangeworks [9], IBM Quantum Experience [10], and QPS [11].[4]

These QC tools, based on a more traditional computer programming paradigm, can still be regarded as inheriting the *punch card* computation paradigm. In it, the user is faced with the clearly delineated step sequence of writing the code, submitting the code to be executed, and waiting for the results to come back. On the other hand, within the CM tools sphere, it is often seen the predominance of a realtime computation paradigm, where the program is being changed as it is being executed.

It is worth noting that, while I offer these simple categories here, painting the landscape with large brushstrokes, there are examples of practices that challenge these broad boundaries. Such is the case with *live coding* [14] where performer-composers can be found programming music live using code-based languages, and often projecting their code on a screen on stage alongside them.

---

[2] From now on, in this chapter, simply referred to as Max.
[3] Most of them are based in the Python programming language.
[4] For a more complete list of current Quantum Computing tools see [12], [13].

Computer Music practice, in its broadest sense, is strongly informed by this realtime computation paradigm. From listening to notes as they are being dropped on a score, to tuning audio effects while a song is playing, and algorithmically generating tones and melodies that respond to a live input on the fly.

## Previous attempts for an integration

Given that only recently QC has become more available to the larger community of researchers and enthusiasts worldwide, in both tools and learning references, the first initiatives to integrate QC with Music Composition mostly came from researchers with a Computer Science background. Unsurprisingly, these attempts have relied heavily on QC code-based tools, meant for non-artistic practice. Such is the case with Hendrik Weimer's *quantenblog*, where he presents some musical examples that were built with his C library for QC simulation, libquantum [15]. As early as 2014, I myself attempted to integrate the (then very obscure) QCL programing language[5] with my compositional practice and electroacoustic setup, with no practical success.

A second generation can be found expressed in the work published by researchers with stronger artistic considerations. The integration strategies present in these works are mostly characterized by more complex systems that include higher software stack requirements, or simply the proposal of a new CM dedicated application altogether. The first generation of the *Quantum Synthesizer* [17], a Max-based synthesizer making use of QC, can illustrate this.

In this first generation of the *Quantum Synthesizer*, a 48 hour hackathon project at the Qiskit Camp Europe, in September 2019 [18], Max is used as a frontend where the user changes selected parameters that are passed to a backend Python environment via OSC.[6] In turn, this Python environment, that can be running on the same machine or somewhere else in the local area network, is configured with Qiskit and running several Python scripts that account for the network information exchange and to programmatically build quantum circuits based on the parameters received from Max. These circuits are then simulated locally (with or without a noise model) or sent to real quantum computer hardware in the cloud. After retrieving the execution results, these are returned to Max, via OSC, which changes the state of the synthesizer accordingly (see figure 2).

[5] The first quantum computing programming language by Bernhard Ömer [16]
[6] Open sound control, a mostly music related, udp-based, networking protocol.

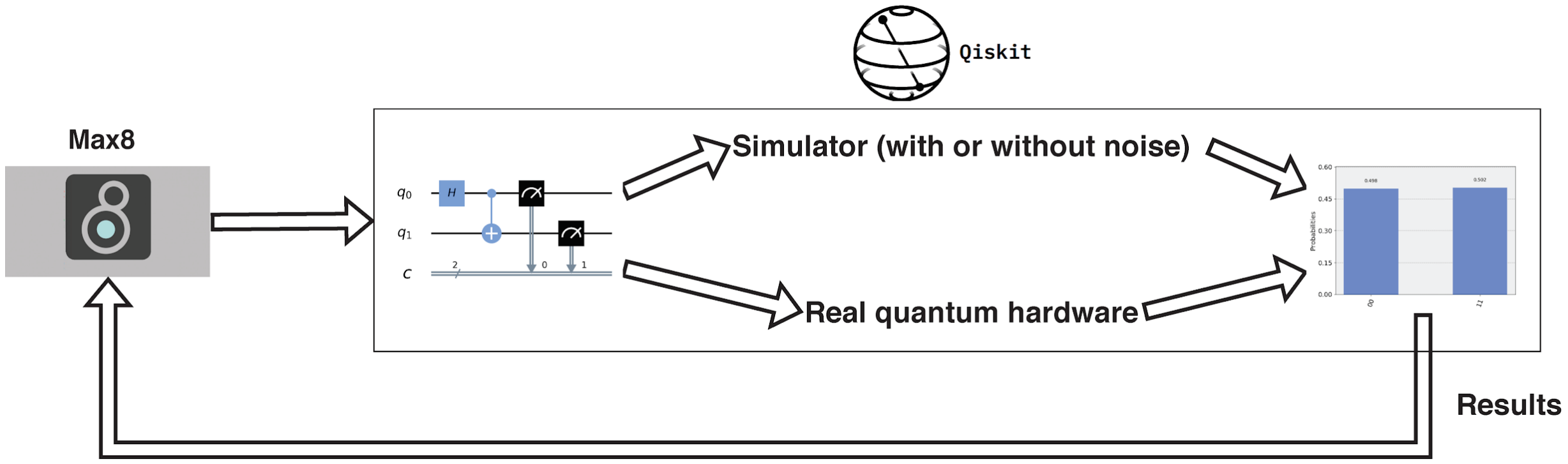

Figure 2 - Architecture of *Quantum Synth*. From [17]

A similar strategy is explored by Eduardo Reck Miranda in his interactive quantum vocal system architecture (see figure 3). In it, there is also a CM system that is connected to a Python environment, within a networked software architecture. However, Miranda's approach relies more heavily on the direct use of Python and Jupyter notebooks, with Csound scripts being triggered from the Python environment [19], [20, p. 17]. The musical output, in this case, is managed through a more code-based interface, which was intended to work more seamlessly with the QC framework. This is at the cost of a higher learning curve, and a *less* realtime CM environment.

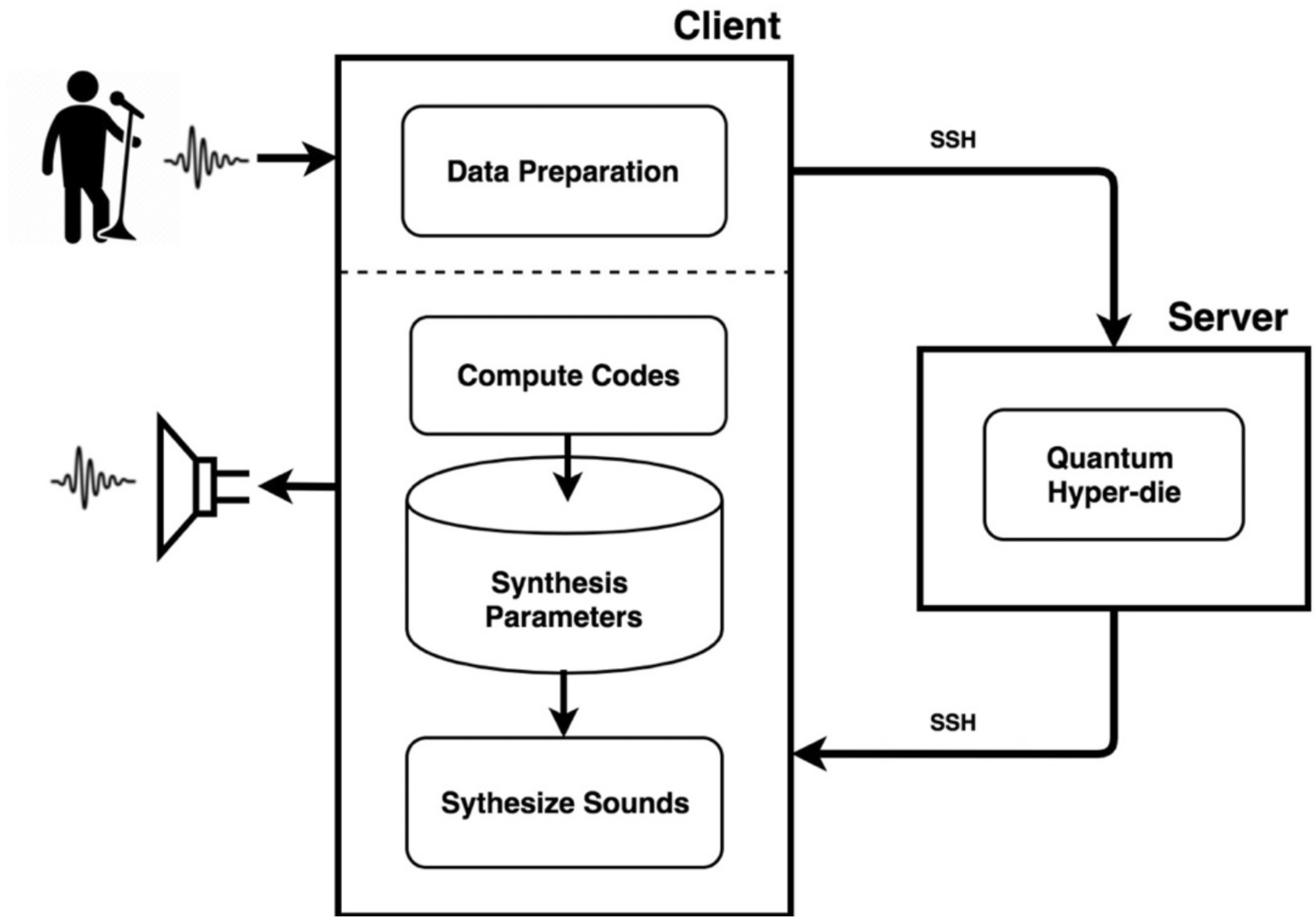

Figure 3 - The interactive quantum vocal system architecture. Reproduced by permission from Eduardo Reck Miranda [20, Fig. 15]

Another approach, taken by James Weaver, has been to create an entirely new CM application environment from scratch. In his web application, *Quantum Music Composer*, Weaver creates a new interface that allows the user to generate 3rd species counterpoint melodies, based on melody and harmony matrices [21]. The app generates output as Lilypond code, that can be rendered as a readable musical score using the Lilypond music score engraver [22]. Though it has a more clean interface, its use is also more restricted than the previous examples.

It is clear from most of these examples that more visual user interfaces, that aren't simply just a terminal window, are more inviting to musicians and creative users. However, it is also clear that most of these implementations still relied on rather complex software infrastructures that are not easy to set up and modify during the creative process. Weaver's system requires considerable non-CM and non-QC skills to modify it and use it to achieve a different compositional goal. Miranda's system requires more knowledge of code-based practices. And my own initial version of the *Quantum Synthesizer*, even in the more inviting user interface that was explored shortly after its hackathon conception (see figure 3), requires different software pieces to be running at the same time and be launched in the correct order.

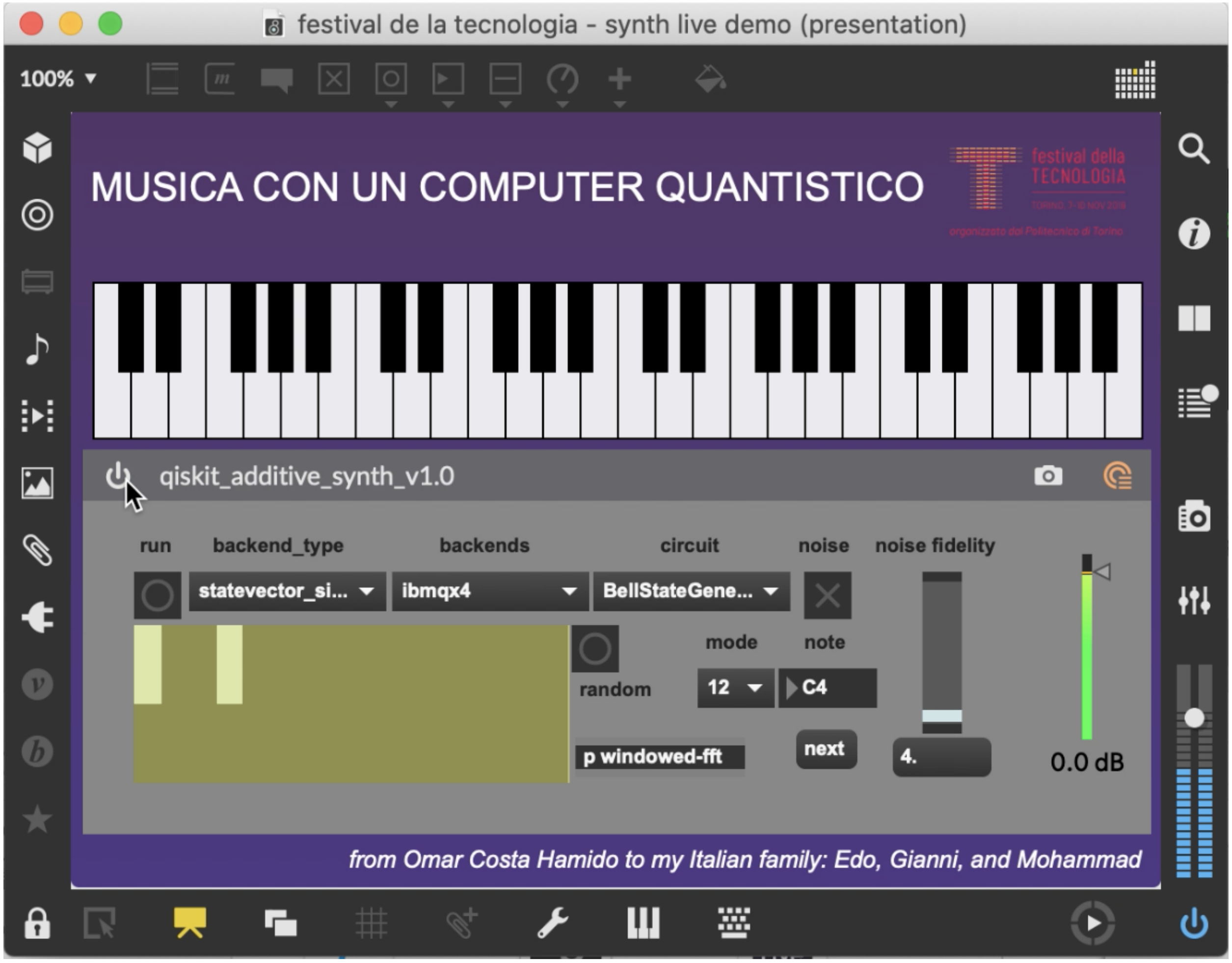

Figure 4 - Second GUI for the Quantum Synth presented in [23]

At this point, there is both a need for more musician friendly interfaces as well as to rethink what is to be expected of this integration and what it should look like. On the one hand it seems that reducing code-based language barriers is one avenue, on the other hand, simplifying the deployment/setup/configuration process of these systems is equally important to make it a more practical tool. For the rest of this chapter, I will give an account of my own work to this effect.

## A new Quantum-computing Aided Composition

When exploring the integration of QC with Music, it is important to have a clear idea of what it is that QC can offer, and set the proper expectations. There are some popular misconceptions about QC that need to be avoided, like the idea of it replacing all the current classical computers when, in reality, as Scott Aaronson and science communication outlets like Quanta articulate, “quantum computers aren’t the next generation of supercomputers - they’re something else entirely” [24]. They are different in the way they function, making use of quantum mechanics phenomena to compute, which requires a different approach to articulate a problem to be computed. QC uses quantum information, but we build and send the set of instructions for the computation (the computing job) using classical machines, and classical information. And, most importantly, what we retrieve from a quantum computer is, in fact, classical information - that is the result of measurement operations that collapse the quantum states into readable classical information (see figure 4).

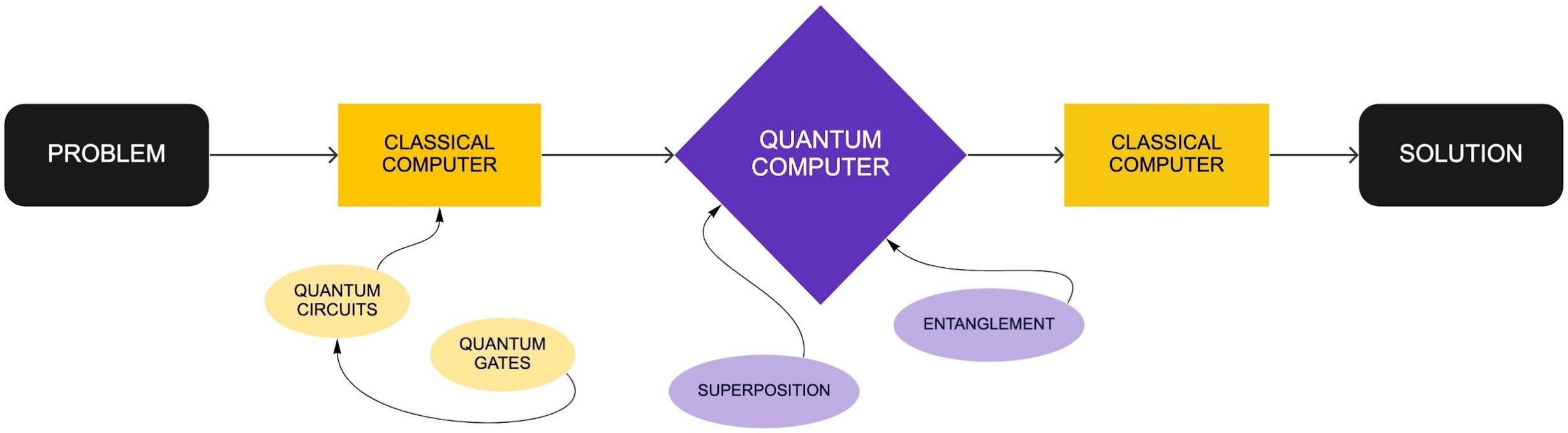

Figure 5 - Overview of computing a problem with QC. From [25]

As music is composed of sound, and sound is the result of a (classical) mechanical force, pulling and pushing air particles around, there will be no “quantum sound” one will be able to hear at some point. QC can enter the world of music in assisting the compositional process, similar to what has been happening with classical computers for the past century. Computer Aided Composition[7] (CAC) has a long tradition, with different approaches to using computers for

[7] Also mentioned in some references as “Computer Assisted Composition”.

sound and music [26], and a new Quantum-computing Aided Composition (QAC) should refer to it.[8]

In effect, creating music that draws analogies with quantum mechanics (QM) is a practice that predates, by several decades, the actual use of QC in music. This calls for a necessary distinction where the former can still be represented by the term “Quantum Music,” adopted by some early creative practitioners,[9] while the latter might be better suited under the term “Quantum Computer Music.” Furthermore, as it will become clear during the rest of the chapter, the QAC concept further articulates a distinction under this umbrella where, as the name implies, QC must support or help the progress of creative practice. For QAC, the creative practice must be the driving force for the integration of QC.

When talking about IBM’s roadmap for QC’s near and long term future, Jay Gambeta refers back to the importance of pursuing QC as a way to tackle seemingly impossible problems “with fresh eyes,” and explains the motivation for their hardware and software improvements. Ending his presentation he states that “[...] if developers can eventually do all this, with a few simple lines of code, using their existing tools, we’ve succeeded.” [29]. This corroborates an important point that I am bringing up in my writing. In order to advance QC, with diverse communities such as the one represented by creative artists, a balance must be found between stimulating thought within a new (computational) paradigm and lifting language barriers. With my work, I propose that using existing creative tools is a way to cross that barrier or, better yet, to more easily expand the current creative vocabulary and stimulate thought in this new paradigm.

A modern-day generation of tools for QAC must emerge from within the CM sphere and allow more seamless integration with music composition workflows. These may have to be simple enough so that the musician won’t have to worry about the underlying software (and hardware) framework. But they also need to be complex enough so that the musician can dive deeper into them, and use it to accomplish different compositional tasks. In practical terms, one might still be tempted to devise 2 approaches at this point. One approach would be to extend current algorithmic music composition libraries in Python, like music21 [30], Abjad [31], or SCAMP [32],[10] which would still require considerable effort in articulating them with the rest of CM tools, and lending them simpler interfaces. The other approach, the one I took, is to make the new tools emerge within an already well known environment and well articulated with the different CM practices, like the Max visual programming environment.

In my research and creative practice, I already use Max quite extensively and I find it to be accessible to musicians and artists working with technology at different levels, and coming from different backgrounds.[11] The ability to quickly create simple interfaces and the strong direct integration with the Ableton Live DAW, makes it the perfect candidate for developing tools for QAC.[12] In the next section I will give an account of the early attempts for the creation of these tools.

---

[8] The term “QAC” was first coined in 2019, in the context of the research found in [27]. It was intentionally not abbreviated to qCAC or QCAC to make it clear that it is not simply an addition to CAC. On the same token, as will become clear in the next paragraph, it was intentionally not translated to simply “Quantum Aided Composition.”

[9] A recent collective manifestation of that can be found in [28]

[10] For more context about these tools see, for example, Tymoczko’s review of music21 [33], and Evanstein’s comparison between algorithmic composition tools in [34].

[11] Another programming environment worth considering for this type of work is Pure Data [35].

## Early attempts for QAC

As it started to become clear for myself what needed to be articulated with QAC and what its contributions to the artistic community would be, it also started to become clear what were the strategies to explore its implementation. It was very important to be able to abstract away all the underlying software infrastructure setup required to simulate quantum circuits, and to eventually communicate with real quantum hardware in the cloud. And all this in a CM environment already familiar to musicians. The creation of a Max for Live device, using the integration of Max in the Ableton Live DAW, available since 2012 [37], was already a clear goal in order to make it available for a larger community of non-programmer electronic musicians. But another integration made available in late 2018, the Node for Max [38], turned out to be equally important in clarifying the path to get there.

In early 2019 I devised several experiments to explore this integration of QC using Node.js, and ended up working very closely with the (now archived) Qiskit.js node module [39]. Even as an experimental module, I was able to incorporate it into Max and successfully deploy a simple Max for Live device called **och.qc-circ-2q**.[13] This was a MIDI device that can be simply drag-and-dropped into a MIDI track chain to add it to the Live session. On the right side of the device the user is presented with a set of 4 sliders that represent the expected probabilities for measuring each of the four possible states that 2 qubits can represent: |00>, |01>, |10>, and |11>. These probabilities change depending on the quantum circuit being used. The left side of the device displays the 3 different ways to load a quantum circuit. The first 2 involve loading Qasm code from either a file on disk or from the clipboard. The third possibility is to read the MIDI clip names in the track (see figure 6).

[12] A similar approach was taken by the Magenta team at Google, where they decided to build Max for Live devices to reach a wider audience of music practitioners [36].

[13] The new Max devices, objects, and abstractions will be highlighted in bold for ease of identification while reading.

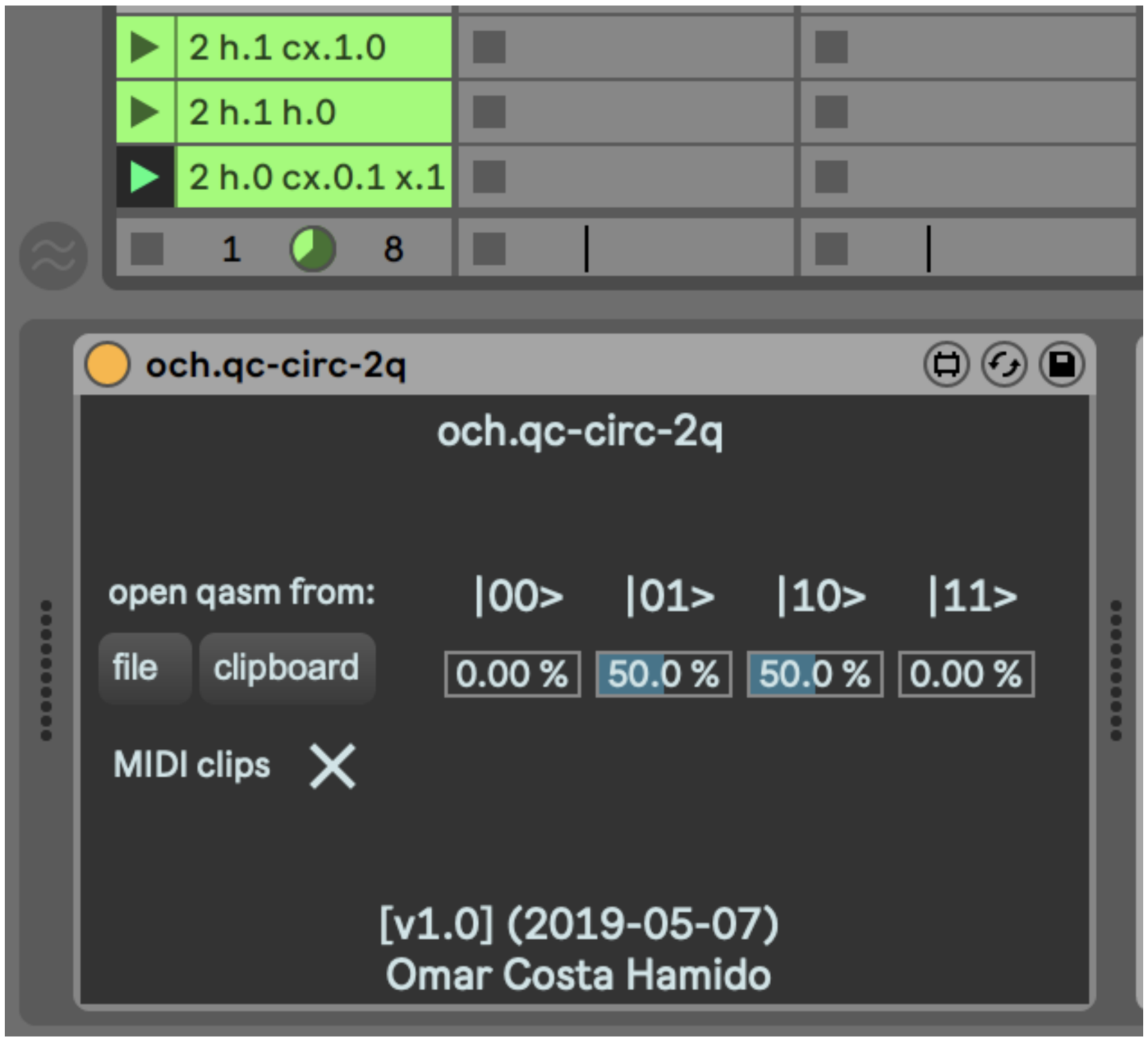

Figure 6 - Earlier generation of fully integrated device. From [27]

Qasm, or OpenQasm, stands for "Open Quantum Assembly Language" and it was proposed as a simple language for describing quantum circuits that can be written or decoded by several of the modern quantum computing frameworks [40]. It is also accepted by different quantum hardware backends, and given its flexibility it was very clear early on that I should articulate a way to integrate Qasm into the device. However, because the generation of the Qasm code can be non-trivial at times, I also decided to implement a new minified circuit notation that I invented for this occasion. This notation, as illustrated in the top portion of figure 6, consists of a list of strings, starting with the number 2 (for 2 qubits), defining which quantum gates to apply to which qubits.

Once the quantum circuit is read, and the probabilities are calculated, displaying the results on the sliders, the user can map each slider value to control other parameters in the Live session.[14] In this case, I used multiple **och.Param2Param** devices to map the slider values to different **och.probGate** devices in several chains of a drum rack. A drum rack in Ableton Live is a virtual instrument that splits incoming MIDI notes to trigger different samples (usually percussive sounds, as part of a drum kit); and **och.probGate** simply serves as a gate that will either block or allow MIDI notes to pass through the chain and play the sample (see figure 7).

[14] Mapping one parameter value to another parameter value is a recurring practice when working with DAWs. Ableton Live comes with a couple devices that allow this, and several more can be found in [41].

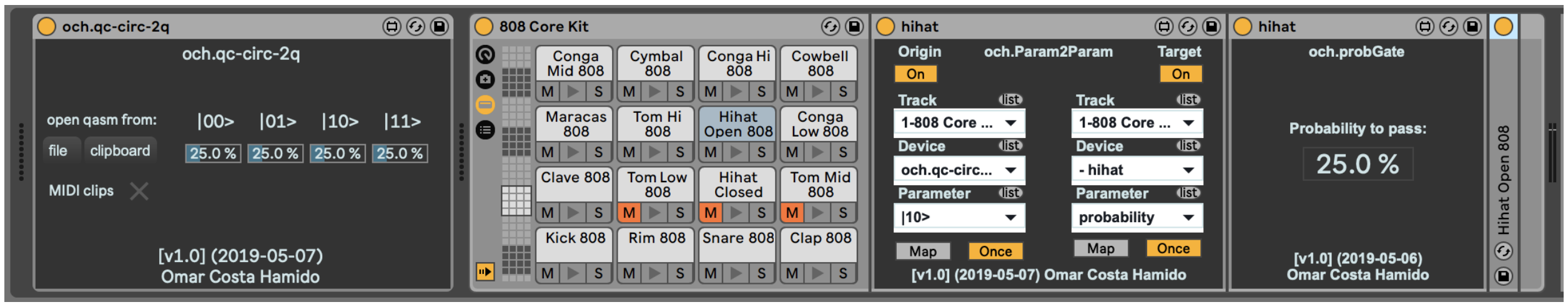

Figure 7 - Example device chain using och.qc-circ-2q together with a drum rack, och.Param2Param, and och.probGate. From [27]

Internally, after receiving the quantum circuit instruction, **och.qc-circ-2q** calculates the quantum statevector[15] and translates that into the predicted probabilities for retrieving each of the possible computational-basis states that 2 qubits can offer. The reason for hardcoding 2 qubits in this system is that performance started to be affected when using a slightly higher number of qubits. With 2 qubits it was still fast enough to be usable in realtime, swapping quantum circuits with no noticeable delay. Some of the factors for this could include the ever increasing library dependencies, sometimes not working perfectly with each other, which made the compilation of this device take more than 10 minutes at some point.

Another important limitation of this device was the fact that it only provided access to the quantum statevector. QC is a probabilistic means of computation, meaning that one only obtains results by running the quantum circuit multiple times and aggregating its results (counts). The distribution of these results should resemble the inner statevector but, by definition, it rarely will be able to express those *clean* values that are calculated mathematically. The actual quantum statevector is, thus, not something that can be retrieved in real quantum computation, but only in simulation. Other factors explaining the variation on the results obtained from real quantum hardware include the noise that disturbs the system and measurement.[16]

The insights gained during these early experiments ultimately meant that I had to abandon this specific approach. Though very important milestones were achieved at this stage while trying to realize QAC, the instability of the underlying framework, a project that was even shut down at some point, as well as the limitations described above, dictated that I had to explore a different approach. It was clear for me that I needed to find a way to implement something much more performant, that would not be limited to an extremely reduced number of qubits, and would still allow the use in a realtime creative practice context.

---

[15] "Statevector" (one word) is in effect a "state vector" (two words), but the former, influenced by its code-based notation, has been largely adopted in the QC world to directly refer to the quantum state, that is, the mathematical description of the internal state of a quantum system.

[16] As explored in [17], there are multiple *outputs* in QC that can be explored: statevector calculation, simulation, simulation with noise models, and real hardware.

## Introducing *The QAC Toolkit*

One of the things that also became clear, after my early attempts for QAC, was that I needed to learn more about how QC frameworks work in order to find the minimum viable implementation that would be less prone to instability. As I started learning more about how the Max environment works, and how to actually write new native objects for this environment, I also became more active in the QC community, participating in hackathons and becoming more involved with the IBM Qiskit community. For some chronological context, in September 2019 I lead a team of Italian engineers creating the aforementioned Quantum Synth [18], by October 2019 I became a Qiskit Advocate [42], in early December 2019 I presented my research to other advocates and IBM researchers, introducing them to QAC and QAD [43], and in mid December 2019 I was invited as a mentor for the Qiskit Camp Africa, in South Africa [44].

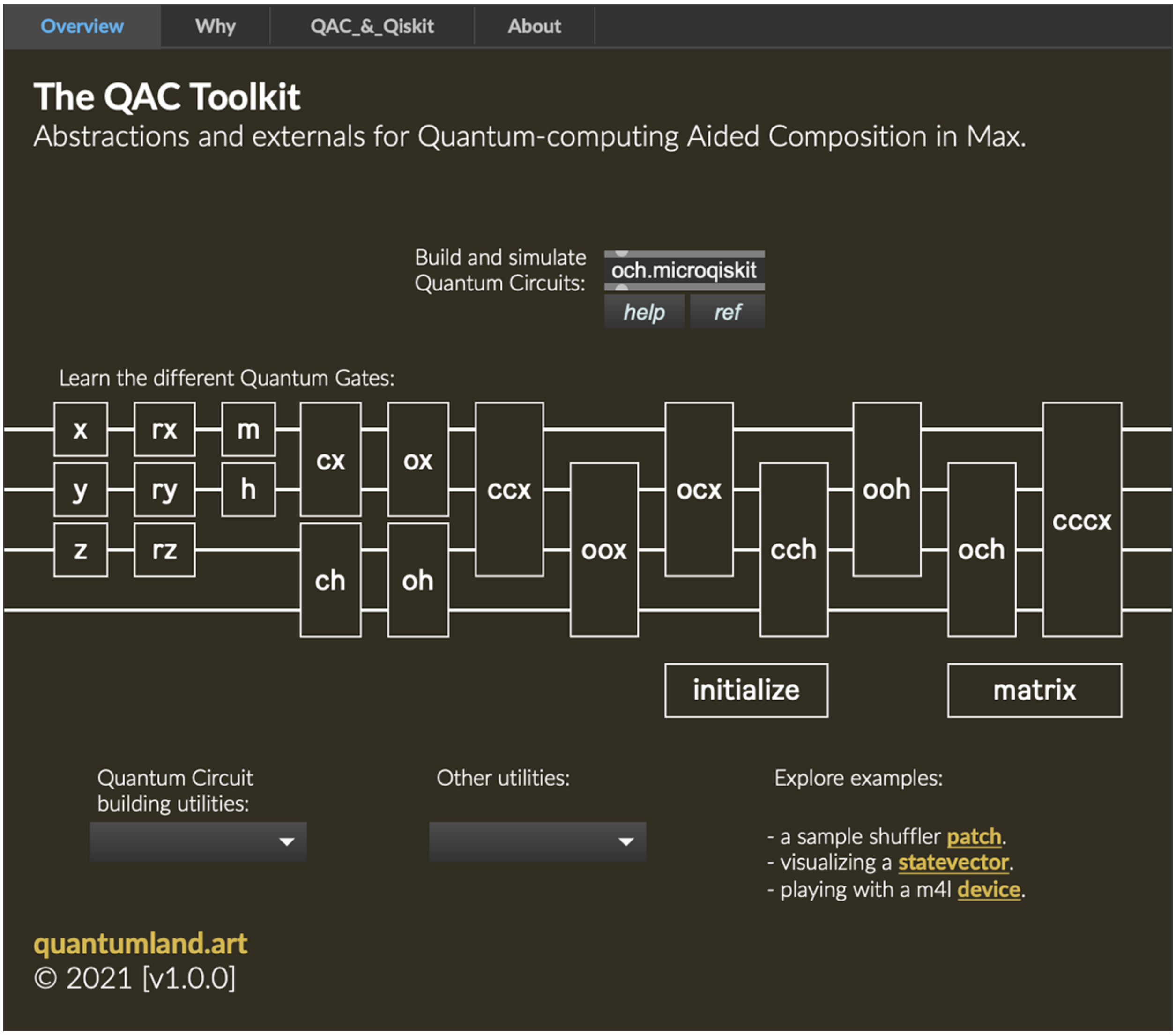

Figure 8 - Screenshot of *The QAC Toolkit* overview patch.

As it will become clear in the next pages of this chapter, this interaction with the community revealed to be very important for the development of my proposed work. By November 2019, I started building my own quantum computing simulator, inspired by the work of Christine Corbett Moran [45], [46] and James Wootton [47]. And between December 2019 and December 2020 I was able to work with James Wootton on creating a new version of the MicroQiskit library in C++ [48] that ultimately enabled me to complete a quantum simulator natively integrated in the Max environment. This new generation of tools, to allow musicians and artists to build, simulate, and run quantum circuits inside Max, gave birth to the software package that I called *The QAC Toolkit* (see figure 8).

*The QAC Toolkit* is a package for Max that is available directly from the Max package manager, and it includes some externals and abstractions for working with QC in Max. This package includes examples for learning about quantum gates and the new workflow directions that these tools enable, as well as several objects to help streamline working with it. One of the main objects, **och.microqiskit**, implements circuit building and simulation functions in Max.[17]

## och.microqiskit

As the name suggests, **och.microqiskit** is very much inspired by the MicroQiskit framework, even sharing several of the method names. However, as a Max object the user interacts with it using Max messages instead of running scripts or using the command line. It should be noted that, besides creating a new object box in the Max patch and typing its name, there is nothing more that is required to set up - no Node modules, no Python packages. This object contains in itself all the required resources to build and simulate valid quantum circuits, as well as to retrieve other information from it.

There are 2 main *entities* (or classes) in **och.microqiskit**, one is the *QuantumCircuit*, representing the definition of a quantum circuit as a group of quantum gates and the number of qubits and classical bits allocated. The other is the *Simulator*, which makes use of the *QuantumCircuit* definition to perform a series of different tasks, including simulation, retrieving the statevector, and converting the circuit definition to other code-based language definitions. All the messages to interact with **och.microqiskit** will always refer to either a *QuantumCircuit* or a *Simulator*. And one instance of this object can hold multiple quantum circuits and simulators.

[17] All the objects start with “och.” because the original short name of the toolkit package is och. This also reflects a recurring practice in object crafting among the Max developer community. Still, some object mappings are available as well: e.g. *och.microqiskit* can also be initialized simply by typing *microqiskit*.

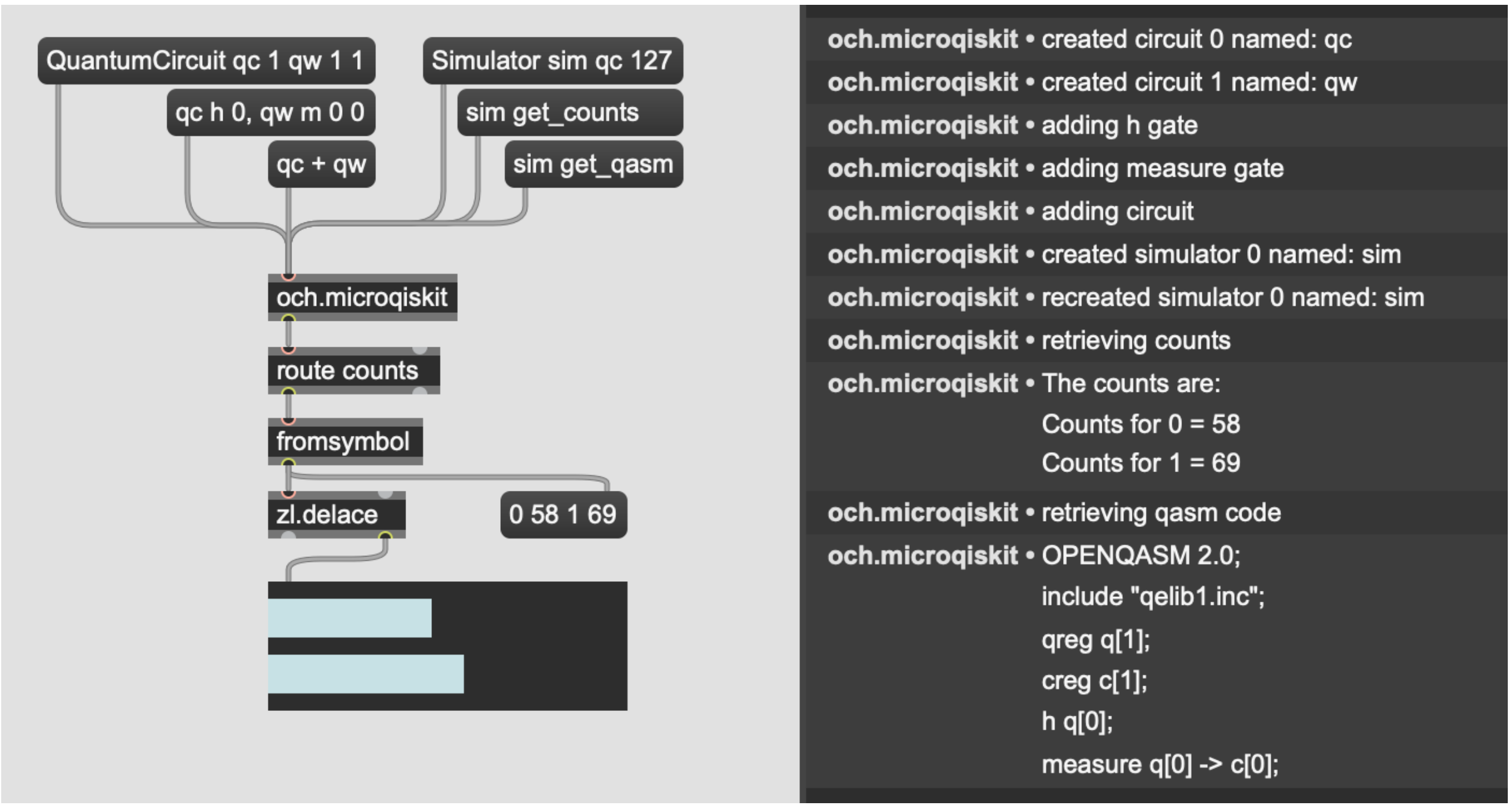

Figure 9 - och.microqiskit object example patch, on the left, and its console output, on the right.

In figure 9 we can see a simple example patch using **och.microqiskit**.[18] On the top portion of the patch there are 2 groups of messages, one for creating and changing the quantum circuit, and another for creating and firing the simulator. The first message "QuantumCircuit qc 1 qw 1 1" creates 2 new *QuantumCircuits* named *qc* and *qw*; the first with 1 qubit, and the second with 1 qubit and 1 classical bit. The second message "qc h 0, qw m 0 0" actually sends 2 consecutive messages,[19] one adds a hadamard gate on qubit 0 to the *qc* circuit, and the other adds a measurement gate between qubit 0 and classical bit 0 on the *qw* circuit. The third message on this left column simply adds the contents of the circuit *qw* to the *qc* circuit.

On the group of messages on the top right, the first message "Simulator sim qc 127" creates a new *Simulator* named *sim* with the contents of the *qc* circuit and sets its configuration to use 127 shots.[20] The second message "sim get_counts" requests a simulation to be run and retrieve the counts of that simulation. In fact, every time this message is triggered a new simulation is executed and new results retrieved. The third message "sim get_qasm" will not run a simulation but will instead convert the quantum circuit into a Qasm definition and output it. By default, **och.microqiskit** will print to the console information regarding each action that was taken (see right side of figure 9),[21] and by doing that the learning process of using it becomes easier. By the same token, it will also print error messages to the console every time there was an incorrect message sent to the object.[22]

---

[18] This Max patch is available for download in [49].

[19] In Max, multiple messages can be grouped in one message box by separating them with a comma. The result is that when the message box is triggered both messages will be sent out individually, one immediately after the other.

[20] One "shot" is the same as one execution of the quantum circuit. As explained in the previous sections, QC needs to execute a quantum circuit multiple times to derive enough information to more faithfully represent the computed solution.

[21] Unless the *console_output* attribute is set to 0. See, for example, how och.microqiskit is used in the next section.

The lower portion of this patch shows an example of using the results retrieved from the quantum simulation to change the state of a multislider object. When asked to retrieve the counts, **och.microqiskit** will output a message that contains the word *counts* followed by a symbol (string) containing the list of states and its respective counts. We use the route object to allow only messages starting with *counts* to pass, and we use the fromsymbol object to transform the symbol back into a list - the contents of the message at this point are shown on the connected message box. The zl.delace object simply splits a list into two lists with every other item, that are output separately on the left and right outlets. At this point the message coming from the right outlet only contains a list of counts (without the states) and this is all that is needed to change the state of 2 sliders in a multislider object: note that the multislider object was configured to to receive values in the range of 0 to 127, which is the number of times that the circuit is being executed.

This is a very simple example of using **och.microqiskit**. Because all the interactions with this object only require using standard objects in the Max environment, it is very easy to integrate it with other workflows in Max. For example, all the messages can be programmatically generated, and may include placeholder values[23] that can interactively change the quantum circuit being built. At the same time, there is some flexibility in the structure of the messages. For example, the user can create new quantum circuits by giving it any single word name,[24] opt for initializing it with only qubits, or with both qubits and classical bits, and even create several circuits in one go, as illustrated in the example above. The **och.microqiskit** object itself can be initialized with some arguments that will already give it a *QuantumCircuit* and *Simulator* already defined.[25]

The *Simulator* memory contains the *QuantumCircuit* definition as it was when it was passed to the simulator, but this can be changed if the *sim_update* attribute is set to 1, which will automatically update the simulator whenever its corresponding quantum circuit changes. And once the simulator is defined, we can retrieve different results like the aggregated counts using *get_counts*, the statevector using *get_statevector*, or each individual shot result using *get_memory*.[26] The resemblance of the notation used with that of Qiskit, MicroQiskit, and some other QC frameworks, is not accidental. This includes the short names used for the quantum gates, like *x* for the NOT gate, *h* for the hadamard gate, *m* for the measurement gate, *cx* for the controlled-NOT gate, etc.[27]

---

[22] For example, if the user tries to call a circuit with a name that has not been previously set on this object instance, it will reply with an error message on the console. The object will also post error messages if, for example, the user tries to add a gate to a circuit on a qubit that is outside of range, tries to run a simulation without measurement gates, or tries to add measurement gates on circuits without classical bits to store them. This helps to make sure that only valid quantum circuit definitions are generated.
[23] For example, in Max, a message box can have the $1 symbol that is replaced by an incoming value.
[24] This is similar to what is already practiced in Max to initialize and refer back to jitter matrices.
[25] As an example, see the patch in figure 13.
[26] For a complete list of methods for the och.microqiskit object, please refer to the help patch and reference page included in the distributed Max package.
[27] See the full list of available gates in the overview patch (reproduced in figure 8) and each individual help patch. General good quick sources for understanding different quantum gates include [50, Ch. 7 Quantum Gates and Circuits], [51], [52]

Given this object running natively in Max, and with all the concerns that went into making it work seamlessly, I devised a test to evaluate how well it performed when compared the original QC tools, in Python - more specifically against Qiskit, and MicroQiskit, in a Python terminal. The experiment consisted of running a very small circuit - 1 qubit and 1 classical bit with 1 hadamard gate followed by 1 measurement gate - for different numbers of shots, in simulation. The experiment was run for 2,000 shots (figure 10), 20,000 shots (figure 11), and 1,000,000 shots, which is the maximum number of shots that Qiskit allows for a single experiment execution (figure 12).

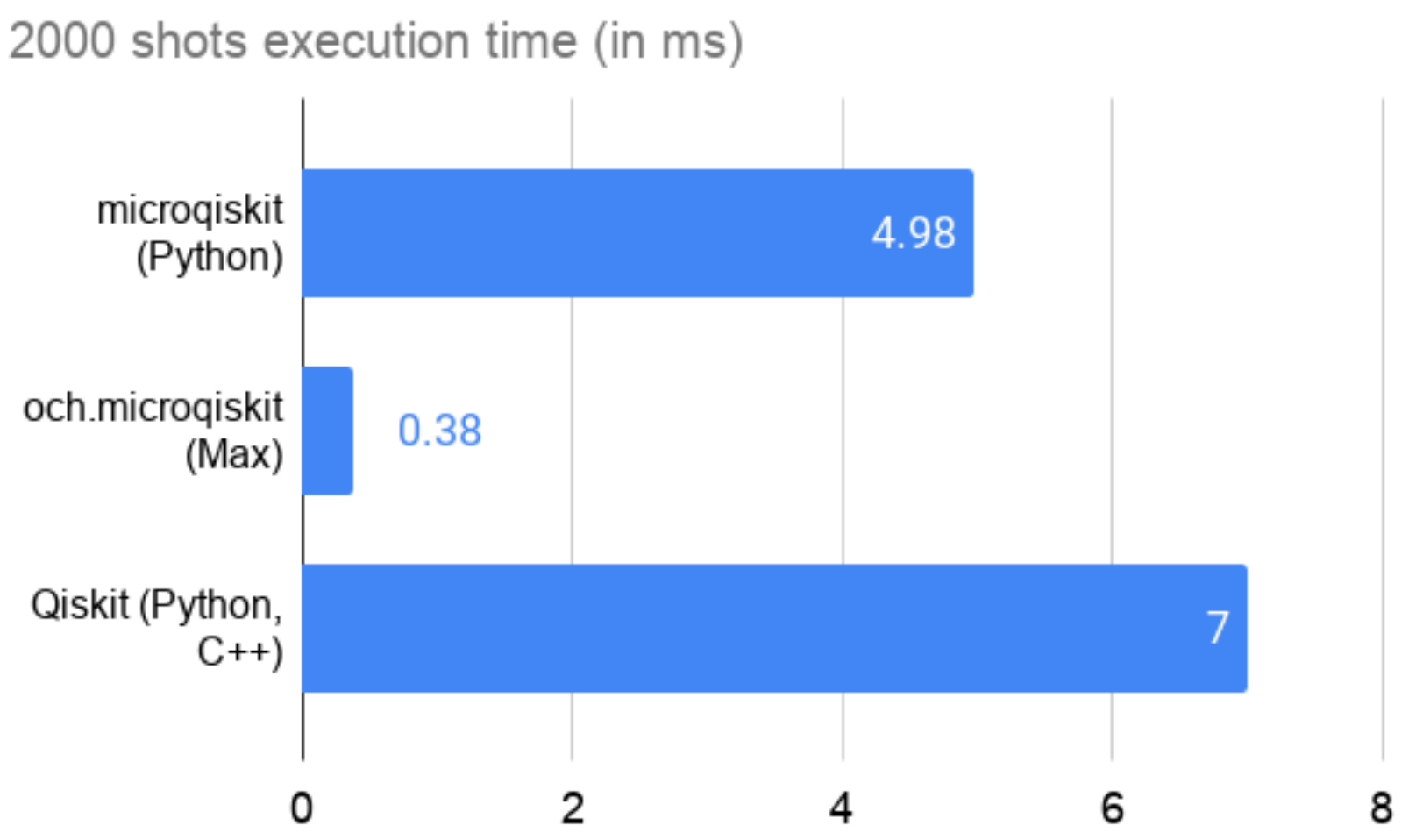

Figure 10 - Comparing MicroQiskit, och.microqiskit, and Qiskit simulating 2,000 shots. From [27]

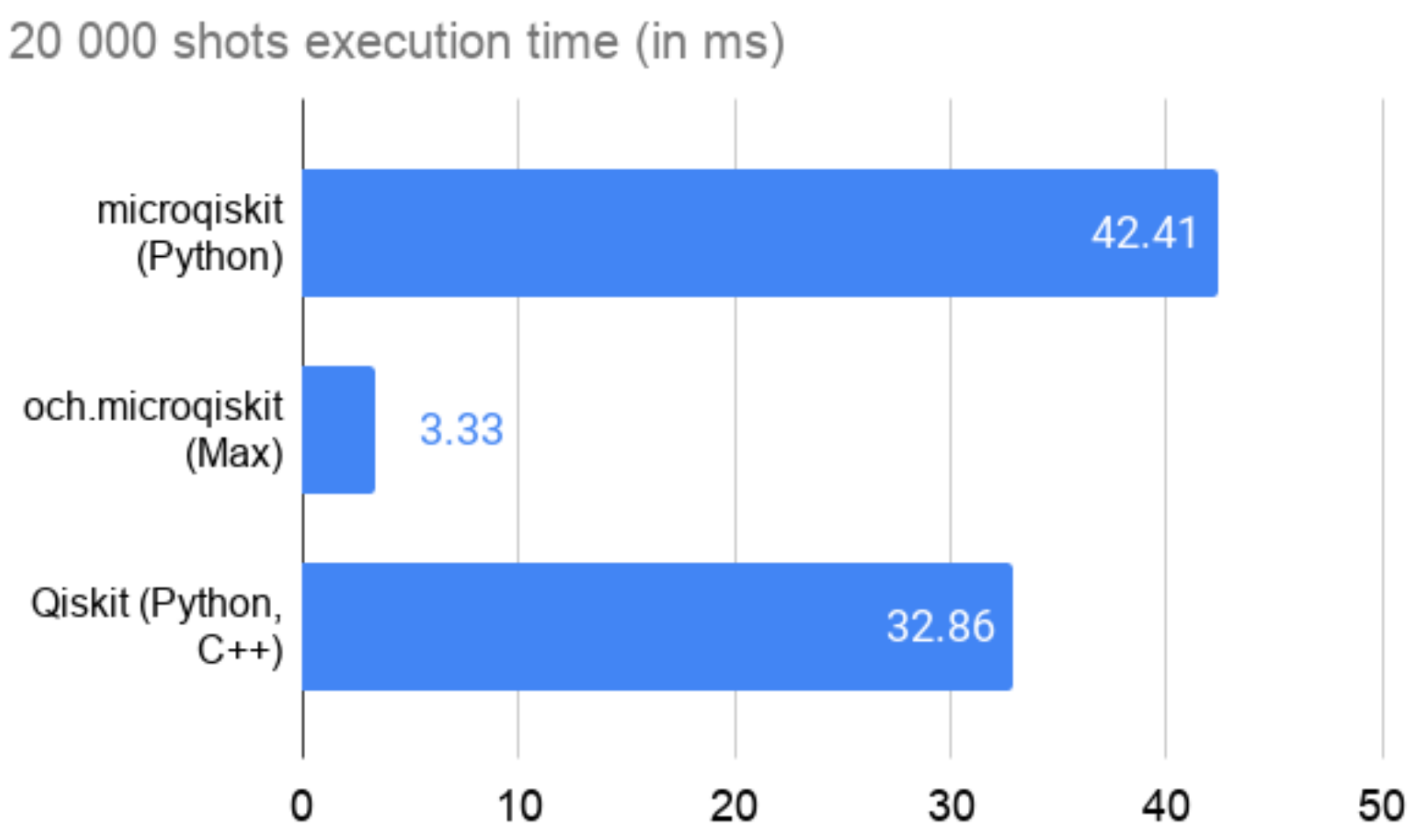

Figure 11 - Comparing MicroQiskit, och.microqiskit, and Qiskit simulating 20,000 shots. From [27]

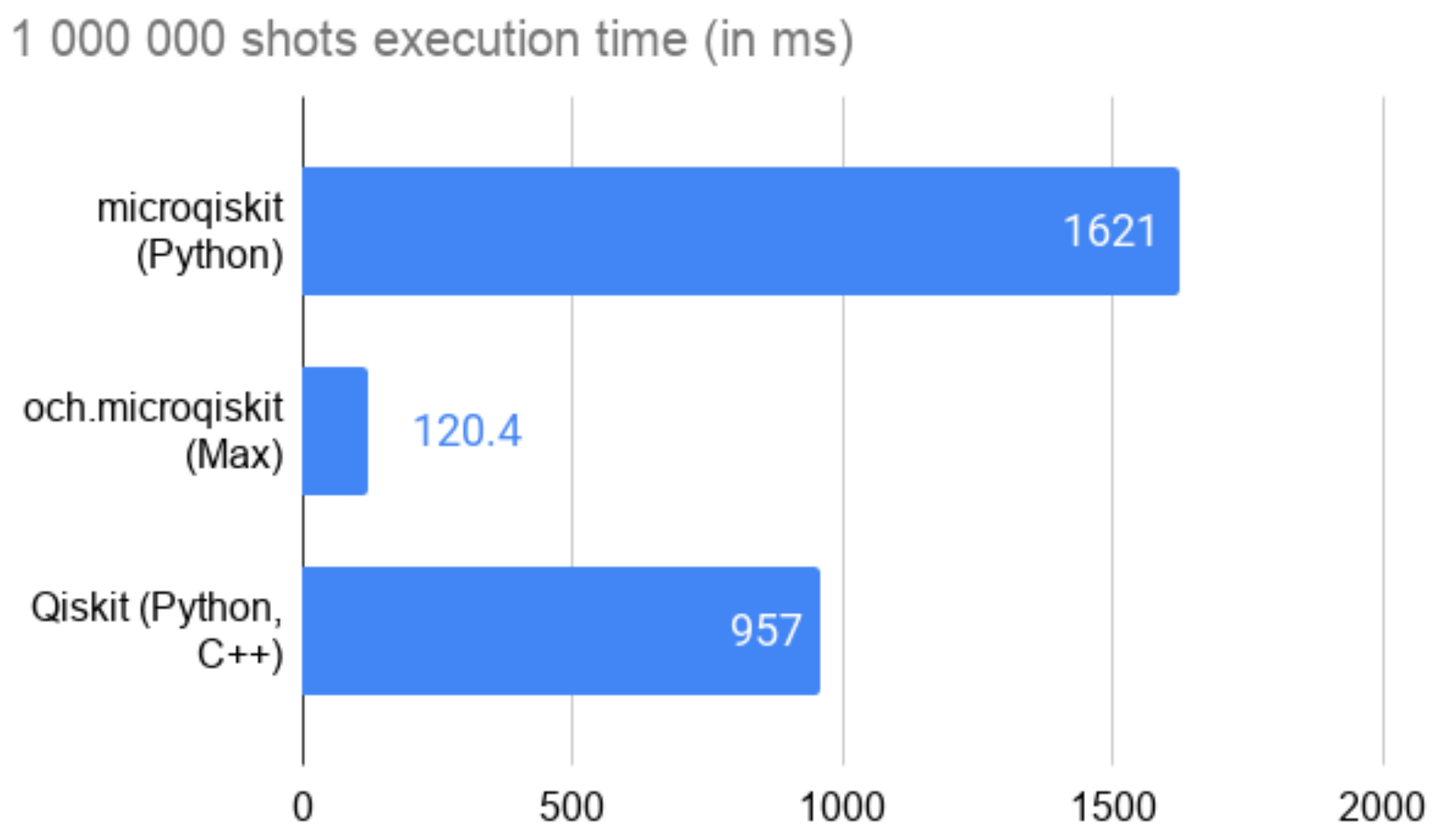

Figure 12 - Comparing MicroQiskit, och.microqiskit, and Qiskit simulating 1,000,000 shots. From [27]

The results, plotted in the figures above, were very surprising. The **och.microqiskit** implementation was consistently 10 times faster than the original MicroQiskit in Python. This can be attributed to the fact that **och.microqiskit** is itself written in C++, which is much more efficient than Python, as has been noted repeatedly for decades [53]–[55]. For 2,000 shots, MicroQiskit is around 40% faster than Qiskit, something that other members in the Qiskit community have also noticed for simple circuits. However, by the time we are executing 20,000 shots, Qiskit is faster than MicroQiskit. This can be due to the fact that Qiskit actually includes some high performance components also written in C++. At 1,000,000 shots Qiskit is about 70% faster than MicroQiskit, and **och.microqiskit** is still 10 times faster than MicroQiskit and 8 times faster than Qiskit.

Being able to simulate QC efficiently is an important step to understand how QC can integrate current creative practices and to learn how to eventually get advantage from the real quantum hardware. As it stands, these tools are already very important for QAC as they are usable in a realtime creative practice context, which was the motivation for pursuing this work in the first place.[28] It should still be noted that it is not possible to prove that computing a problem using QC simulation will be necessarily faster than any other strategy for computation using classical machines. If we consider this idea, finding ways to connect with real quantum hardware will still be an important avenue to pursue, in order to allow creative artists to be ready to benefit from the quantum advantage in the (hopefully near) future.

## och.qisjob

This possibility to connect to real quantum hardware was impossible to achieve with the tools and resources used in the early attempts for QAC (see previous section), and even the original MicroQiskit framework has that limitation. By extension, **och.microqiskit** also doesn't include in itself the resources to establish that direct connection. However, it does have some features in place that are important to achieve it. In order to address this challenge of connecting with real quantum hardware in the cloud I've considered recreating the Qiskit library components that allow the remote connection, but the amount of necessary dependencies make it a very arduous task. Proof of that is the fact that, up until the time of this writing, I haven't met anyone else that would be interested in doing it too.

[28] For more about the discussion on the advantages of QC for the creative practices, please see [27, Ch. 3: QAD and more].

Connecting to full Qiskit (in Python) in some way seemed unavoidable. During my explorations, I've experimented with very old Max objects that should be able to interface with Python directly, but most of them would either only work with older versions of Python, which are not supported by the modern QC frameworks, or they would still need the user to have Python configured on his/her machine anyway. Other strategies considered included deploying a web server with a Qiskit configuration, to relay the jobs to the real machines,[29] but concerns about scalability and managing potential sensitive access credentials cautioned me not to pursue this path on my own.

I've, once again, turned to the community in search of ideas for this seemingly small problem of simplifying the process of submitting a job to real quantum hardware. It so happened that my former hackathon partner Jack Woehr already had a project that he also started in early 2019 that addressed just that: a simpler command line interface for submitting jobs to quantum computing hardware [57]. A job is simply the set of instructions to run on the quantum computer, which include the quantum circuit definition, the number of shots, and some other required information that is compiled by QC framework in order to successfully submit it. Jack Woehr's QisJob is a program that only requires the user to pass an OpenQasm file (or a Python script file with a circuit defined in Qiskit) and automatically executes all the necessary Python and Qiskit functions to successfully submit the job. In effect, it is able to do so because it includes a version of Qiskit as a module in its code.

Between late 2020 and early 2021 I contributed to Jack's project while, at the same time, started exploring the possibility of compiling it into a single application that could connect directly with the Max environment. The **och.qisjob** object emerged from this exploration.[30] This is a two-part object where it includes an OS-specific application, running in the background, and a max object that automatically starts that application and connects directly to it. This approach still removes the need for creative artists to manually set up a Python environment and establish the necessary connections between both environments. Furthermore, this approach also works well with the previous objects and what they already offer.

[29] A process similar to one I explored in my QC-enabled interactive story in 2019 [56].
[30] Note that, as of this writing, och.qisjob is still experimental, and only available in beta.

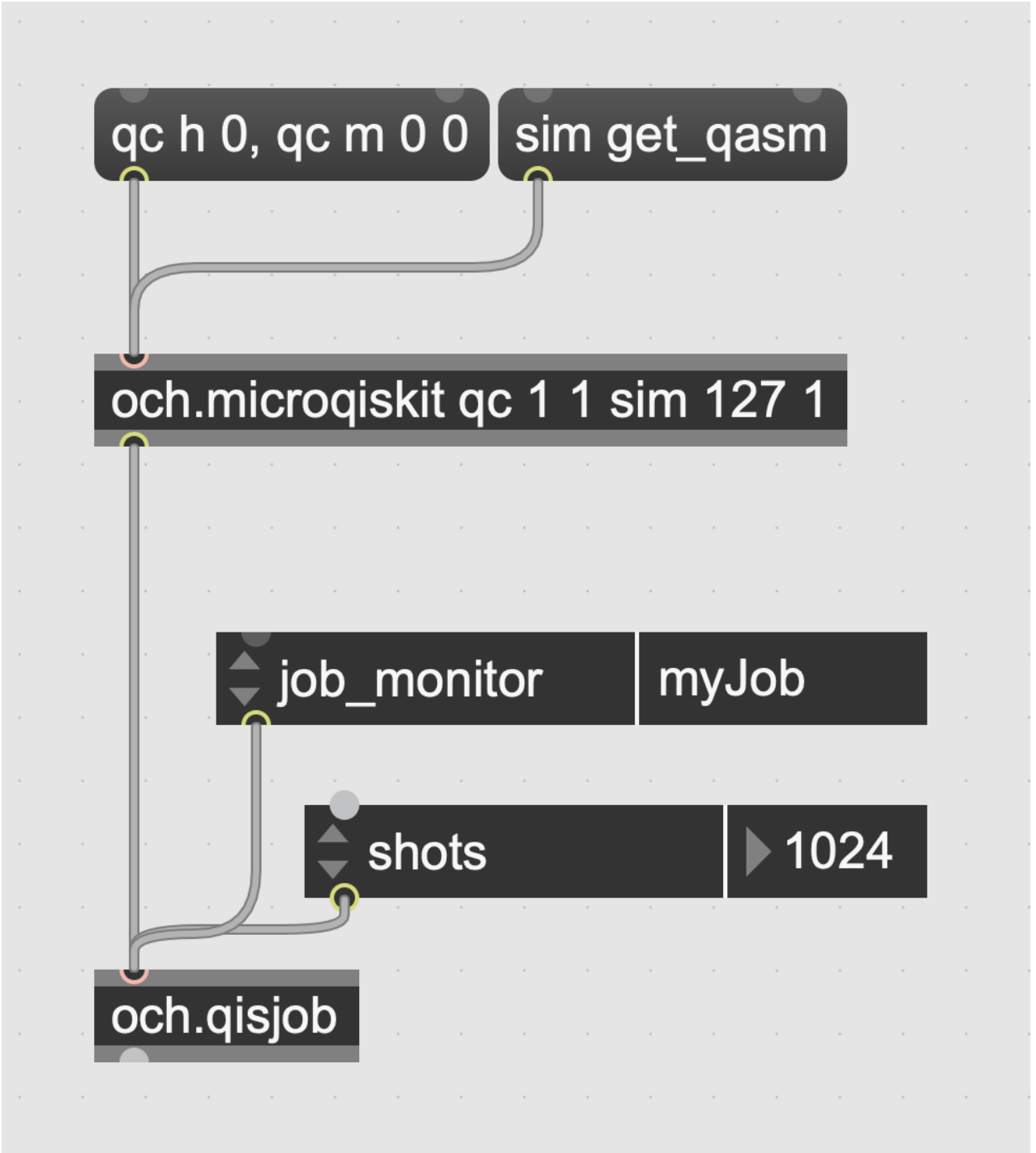

Figure 13 - och.qisjob object example patch. From [27]

The **och.microqiskit** object already offers the possibility to retrieve the Qiskit or Qasm code definitions for a quantum circuit - using the *get_qiskit* and *get_qasm* methods, respectively.[31] This is very useful when working with **och.qisjob** since we can take advantage of the circuit building capabilities introduced earlier. Figure 13 shows an example of what it looks like to work with this new object. On the top portion of this patch the **och.microqiskit** object is initialized with some arguments that define a *QuantumCircuit* named *qc*, with 1 qubit and 1 classical bit, as well as a *Simulator* named *sim*, using 127 shots and with auto update enabled (also known as the *sim_update* attribute). The first message “qc h 0, qc m 0 0” adds an hadamard gate to the *qc* circuit on qubit 0, and also adds a measurement gate between qubit 0 and classical bit 0. The “sim get_qasm” message requests the *Simulator sim*, holding the *QuantumCircuit qc*, to translate the circuit definition into Qasm code.

The Qasm code generated is sent directly to the **och.qisjob** object that, in this example, also shows 2 attrui objects connected that allow both inspecting and changing the *job_monitor* and *shots* attributes. As an object whose sole purpose is to manage and submit jobs to different backends and retrieve its results, it can also offer the possibility to keep track of the progress of a job using a job monitor function that is offered by Qiskit. The *job_monitor* attribute simply defines the name of the job monitor that will appear in the console window, when it starts reporting back on the job. This is especially useful since most of the publicly available quantum hardware machines use queues to manage all the incoming jobs; with the job monitor the user can know on which place in the queue the job is, when it is being executed, and when it is reading the results.

[31] Internally, och.microqiskit does not use qiskit or qasm code to simulate the quantum circuit. Calling these methods triggers a translation of the internal quantum circuit memory. Furthermore, when simulating, the internal representation of the quantum system is a statevector; calling the *get_statevector* method can retrieve the most direct reading of this object’s inner computed state.

There are other messages to interact with the **och.qisjob** object to determine which backends to use, as well as to load the necessary credentials to access them. For the purpose of this chapter however, instead of providing a very detailed description of all the functionalities of this object, which is still under development, I will present the performance measurements and some insights obtained from it. Given that this object appears as a response to the previous strategies described in the section "Previous attempts for an integration," the most direct comparison will be between this object and a Qiskit Python environment connected to Max via OSC.[32]

While devising the test to evaluate the performance of these two strategies, it became clear that there was no way to guarantee that each iteration would spend the exact same amount of time in the queue, when waiting to be executed by the real quantum hardware. For the sake of consistency, I ended up performing these tests using the high performance simulator found in Qiskit that is accessible to both, as a type of backend to request. Using a similar circuit the one presented above, this experiment was run for 2,000 shots (figure 14), then 20,000 shots (figure 15), and finally 1,000,000 shots (figure 16).

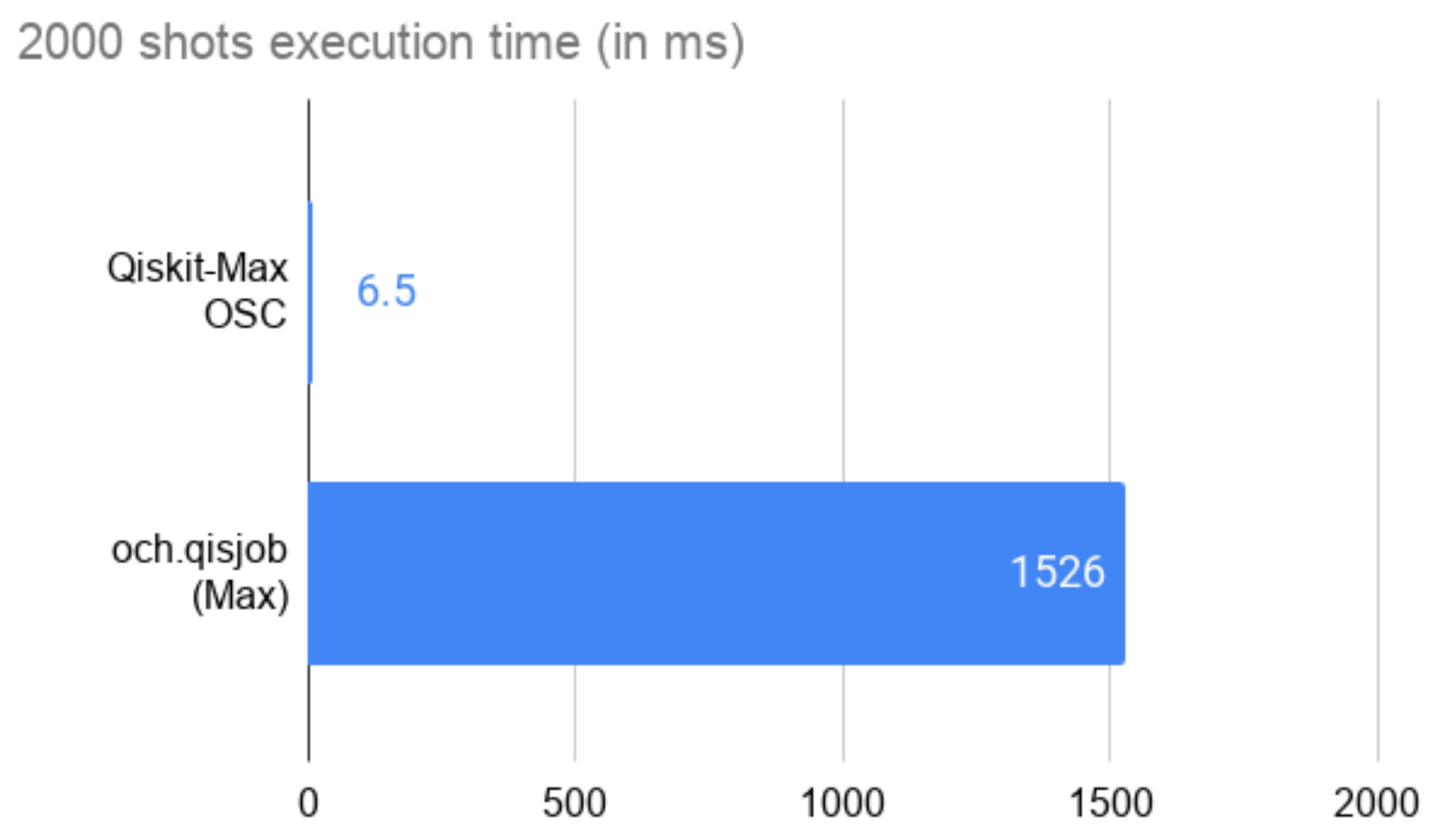

Figure 14 - Comparing Qiskit-Max OSC architecture and och.qisjob simulating 2,000 shots. From [27]

[32] For simplicity, it will be referred to as "Qiskit-Max OSC" in the next pages. It consists of a Python environment with both Qiskit and OSC installed where, upon receiving a message from Max via OSC, it will trigger the circuit execution and then return the results to Max, via OSC.

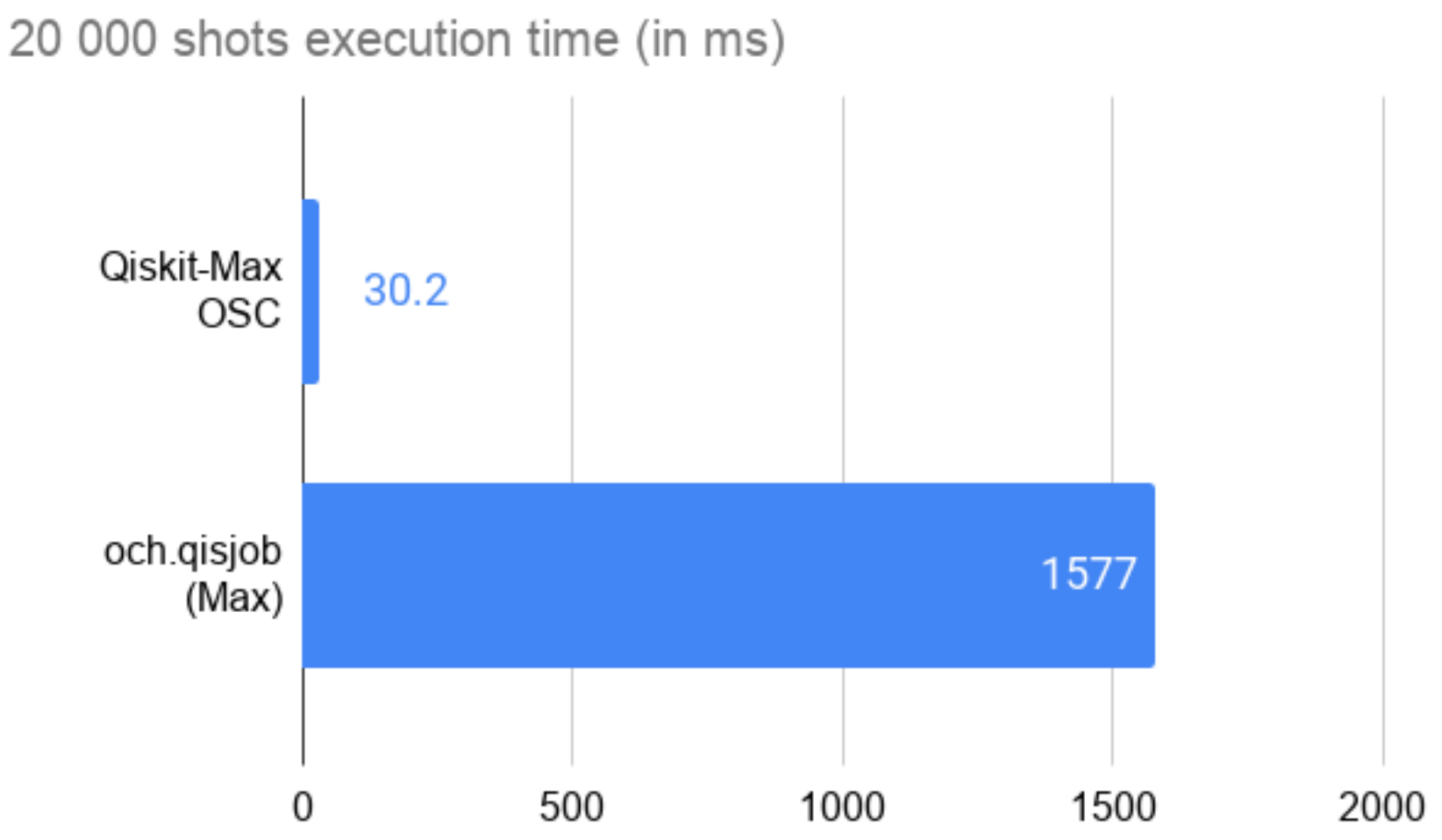

Figure 15 - Comparing Qiskit-Max OSC architecture and och.qisjob simulating 20,000 shots. From [27]

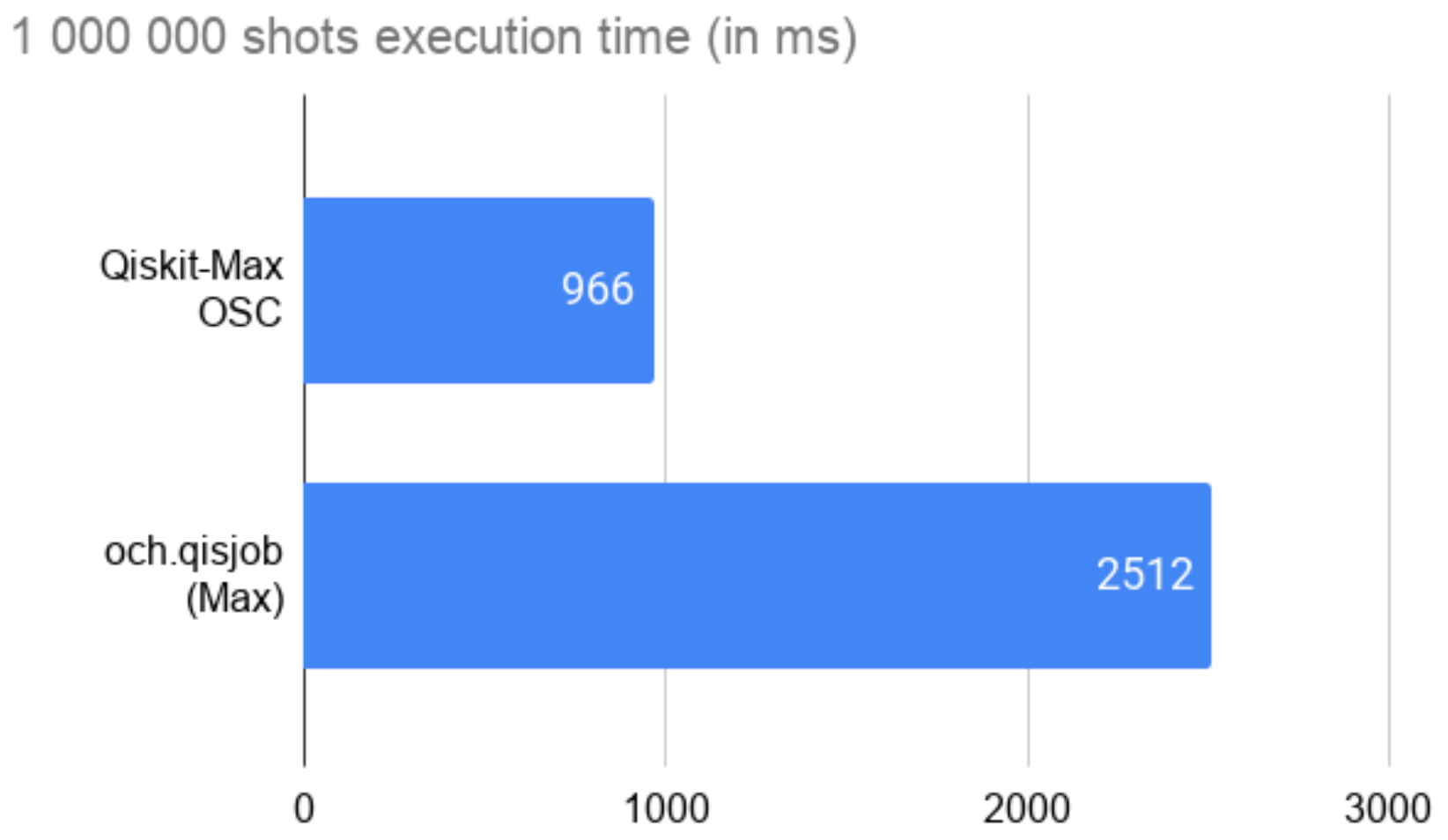

Figure 16 - Comparing Qiskit-Max OSC architecture and och.qisjob simulating 1,000,000 shots. From [27]

The results, plotted in the figures above, were once again very surprising. Even though both strategies ultimately used the same underlying backend system, the high performance simulator provided by Qiskit, their execution times differ greatly. The **och.qisjob** object was consistently worse than the Qiskit-Max OSC strategy. A closer inspection of these results reveals the additional execution time for running the experiment with a higher number of shots increases in the same exact proportion. Somewhere along the way, the current implementation of **och.qisjob** is spending roughly 1.5 seconds every time it is activated.

Debugging so far has suggested that the major bottleneck may be directly related to the fact that, with each new job, **och.qisjob** launches the compiled QisJob application every time, to run in the background, something that takes time given the amount of resources that need to be loaded. This does not happen in the Qiskit-Max OSC strategy explored, since the Python environment needs to be continuously running prior to start receiving the OSC messages that trigger the backend simulation.

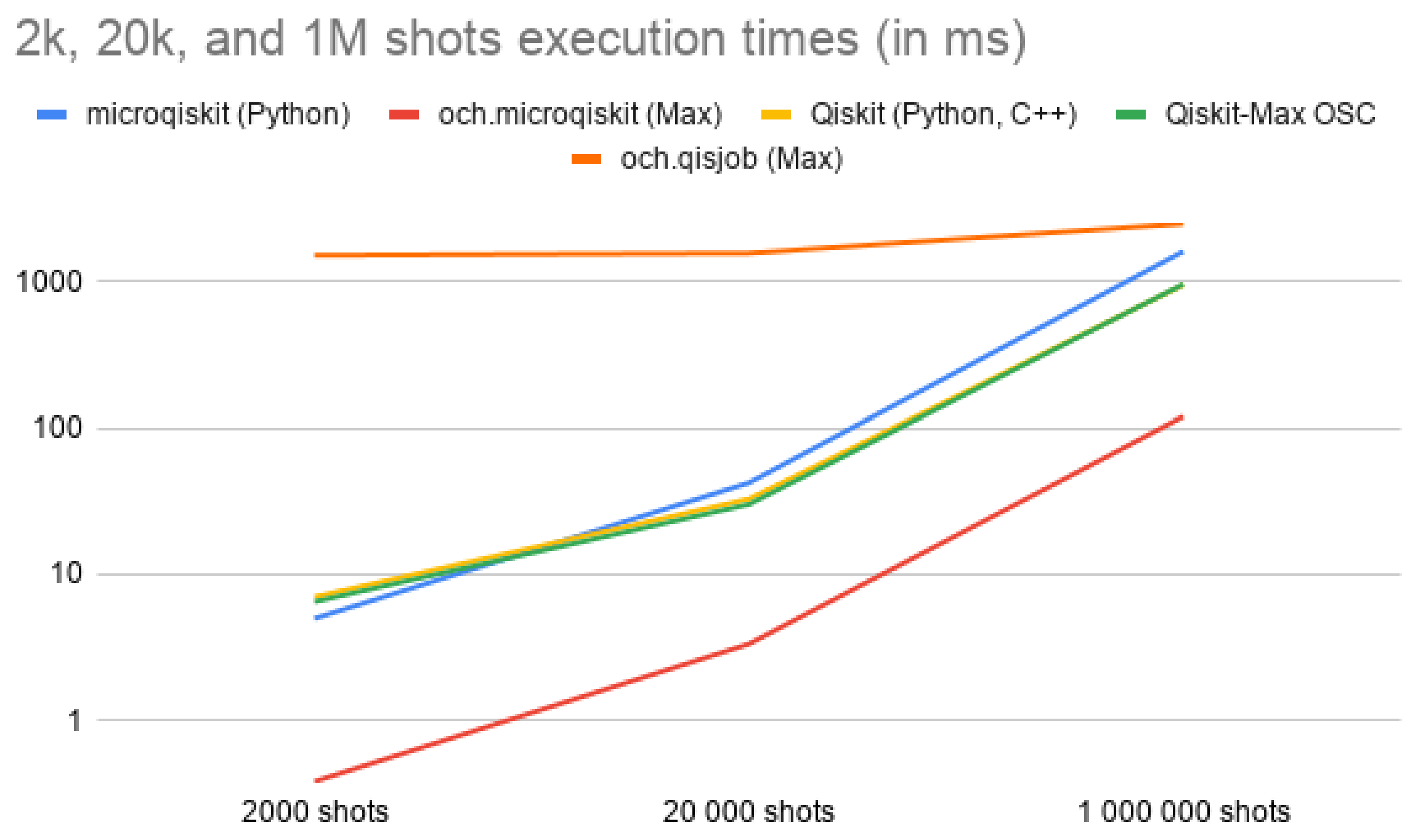

Figure 17 - All execution times for all the tools and architectures being compared. From [27]

Figure 17 combines all the results retrieved from the different tests run. It becomes evident that the Qiskit-Max OSC strategy returned practically identical results to only running Qiskit in the Python terminal, as their lines overlap, which was to be expected since their setups were very similar. It also becomes clear that, among the implementation strategies evaluated here, **och.microqiskit** runs faster than anything else, and **och.qisjob** runs slower than anything else. Further work will be pursued to reduce the execution time of **och.qisjob**. And in the meantime, creative artists that are familiar with working with Python may want to consider using a project like OSC-Qasm: in effect a Qiskit-Max OSC system, inspired by QisJob and **och.qisjob**, offering a python module, a max abstraction, and respective documentation, as the minimum viable product for setting up this type of system [58].

In any case, it is important to consider that, when working with real quantum hardware, most users will be faced with a fair-share queue access system, as described, for example, in [59]. Keeping that in mind, the current 1.5 seconds of execution time overhead in **och.qisjob** might not be that concerning in a real use case scenario. Jobs submitted to be executed on real quantum hardware will always have to wait in a queue, somewhere between several seconds to several minutes.[33] For what they were meant to be, the tools presented in this chapter were very successful in addressing the main concerns. They emerge from within the CM sphere to allow creative artists to integrate QC in their practice, without the need to set up additional environments or be forced to take a traditional code-based approach. With the development of these tools being informed by realtime creative practices, it is equally important the fact that they are fast enough to be used in that context (even more so than some of the alternatives).

[33] There might be providers that can offer to schedule dedicated access time to a machine that can reduce the roundtrip time, but for the majority of users their jobs will have to sit in some sort of a queue for a shared machine. One can anticipate a major revolution when “personal quantum computers” will become widely available.

Going back to the two conceptual spheres for CM and QC, articulated earlier in this chapter, we can now visualize the place of the different systems and integration strategies in it (see figure 18). The same way Qiskit and MicroQiskit inhabit the QC tools sphere, **och.microqiskit** and **och.qisjob** will inhabit the CM tools sphere. As for the previous attempts for an integration, which include works by myself, Miranda, and Weaver, they stand somewhere in the middle with one part in each sphere, as well as a portion outside of both, representing the required work that is not directly related to neither practice. Even though it still needs some work ahead, **och.qisjob** seems to be unique in pointing to a suitable direction to offer a direct connection between these two spheres. Considering this more conceptual and purely technical presentation of how QAC is currently implemented by *The QAC Toolkit*, I will present next a more thorough practical example, showing how a quantum computing algorithm can be implemented.

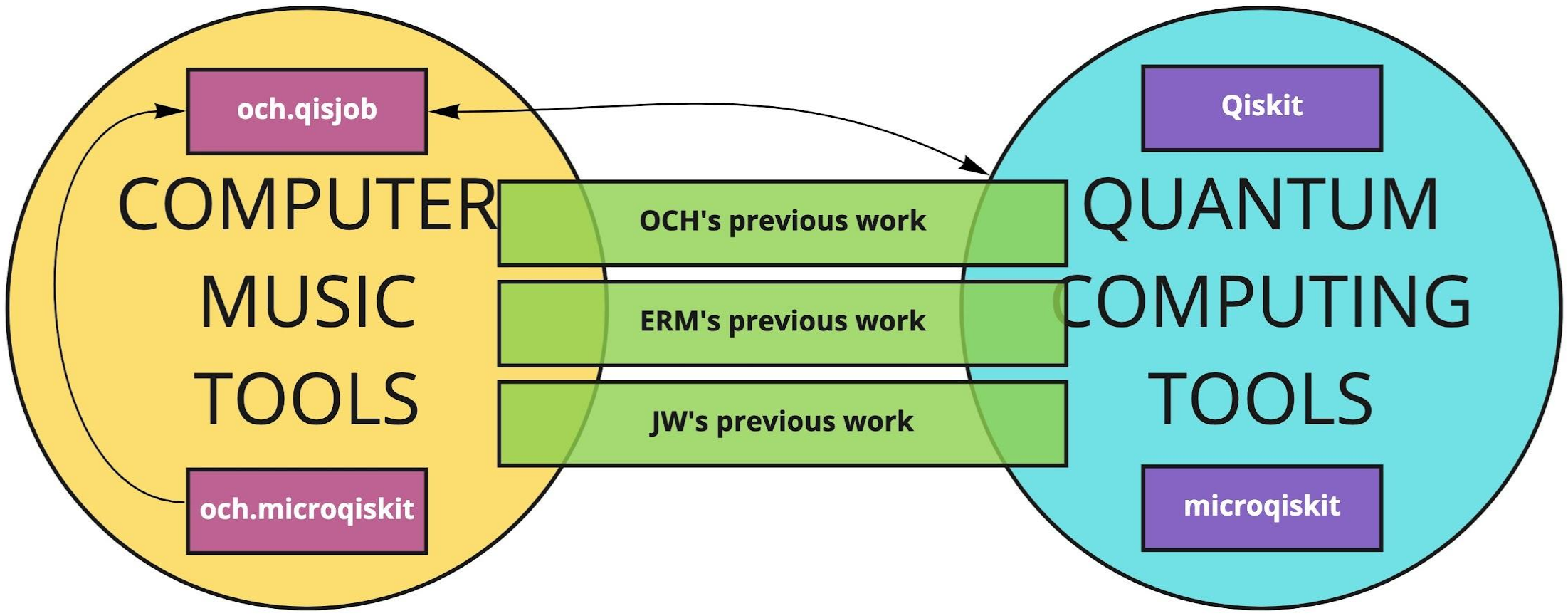

Figure 18 - Computer Music Tools and Quantum Computing Tools, with some specific tools being compared, in context. From [27]

## Implementing BMA with The QAC Toolkit

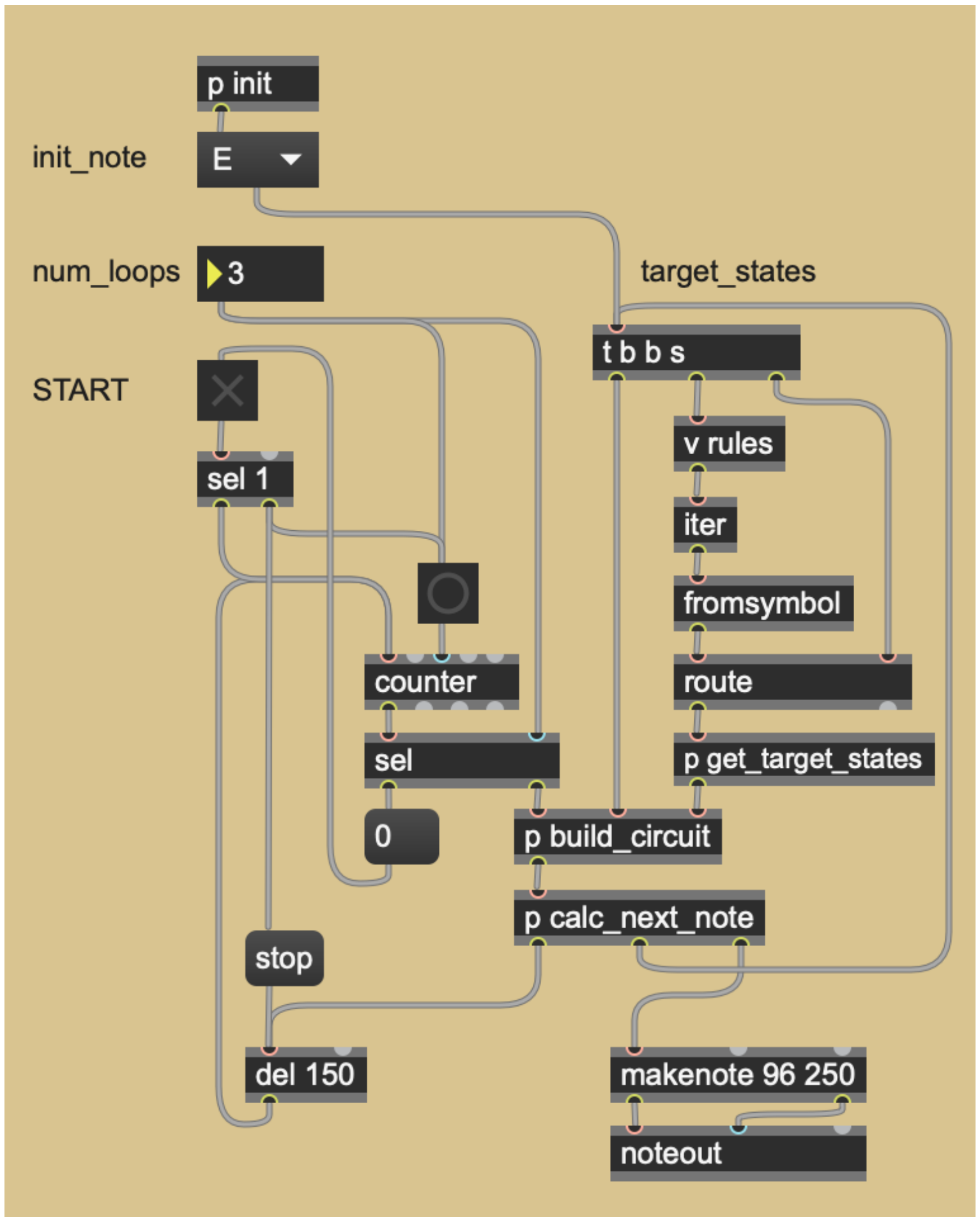

Figure 19 - The Basak-Miranda algorithm running natively in Max with *The QAC Toolkit*

In this section I will dive deeper into using *The QAC Toolkit* by exploring my implementation of the Basak-Miranda algorithm.[34] The BMA is an algorithm proposed by Suchitra Basak and Eduardo Reck Miranda, where they create a first-order markov chain pitch sequencing system making use of quantum computing. Given a pitch transition table, each iteration of the algorithm will extract a row from it and modify an identity matrix based on the possible target pitches, which are translated here into quantum states. The state with highest number of counts, upon execution and measurement of the quantum circuit, is then translated back into the MIDI pitch domain, played on a MIDI instrument device, and used to retrieve the next row of the transition table [60].

Together with their theoretical explanation, Basak and Miranda also released Jupyter notebooks containing the Python code to implement BMA using Qiskit [61]. The rest of this section follows some of the strategies in that example code, sharing function names for some clarity, whenever possible. In figure 19 we can see the main Max patch that displays the overall structure of this implementation as well as the few objects that the user interacts with.[35]

[34] From now on referred to as BMA.

On the top left of the main patch, the 3 user interface objects allow selecting the initial pitch, the number of iterations to run, and toggle the start of the process, with the *umenu*, *numbox*, and *toggle* box, respectively. The *init* patcher object above the *init_note* menu contains the complete transition table and labels, as defined in the example provided by Basak and Miranda (see figure 20). These are stored in 2 value objects, *next_notes* and *rules*, which are accessed remotely from other sections of the patch. In effect, as soon as the user selects a pitch from the *umenu* box, the patch will retrieve all the *rules*, from a value object, and route the one that starts with the selected pitch name to the *get_target_states* patcher.

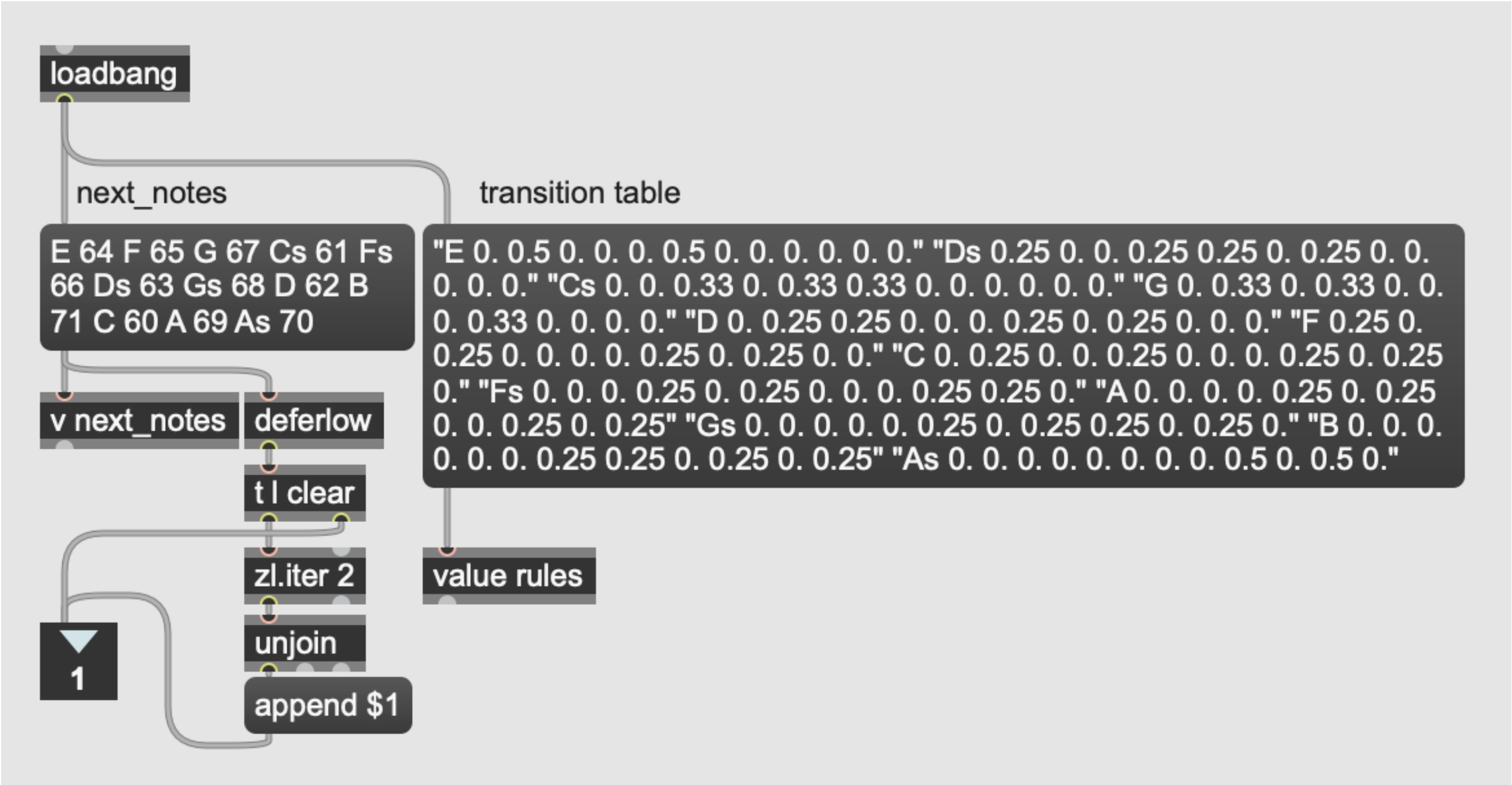

Figure 20 - Contents of *init* patcher

The *get_target_states* patcher simply transforms every non-zero value from the selected transition table row into a value of 1 (see figure 21). This tiny patch includes another subpatcher of its own, named *warn*, that simply serves as an aid to let the user know that BMA, as initially described,[36] will work as intended only when the total number of target states is less than half of the total number of possible states that the current number of qubits can represent (see figure 22).

[35] This patch is available for download in [49].
[36] As of the pre-print available in [60] at the time of this writing.

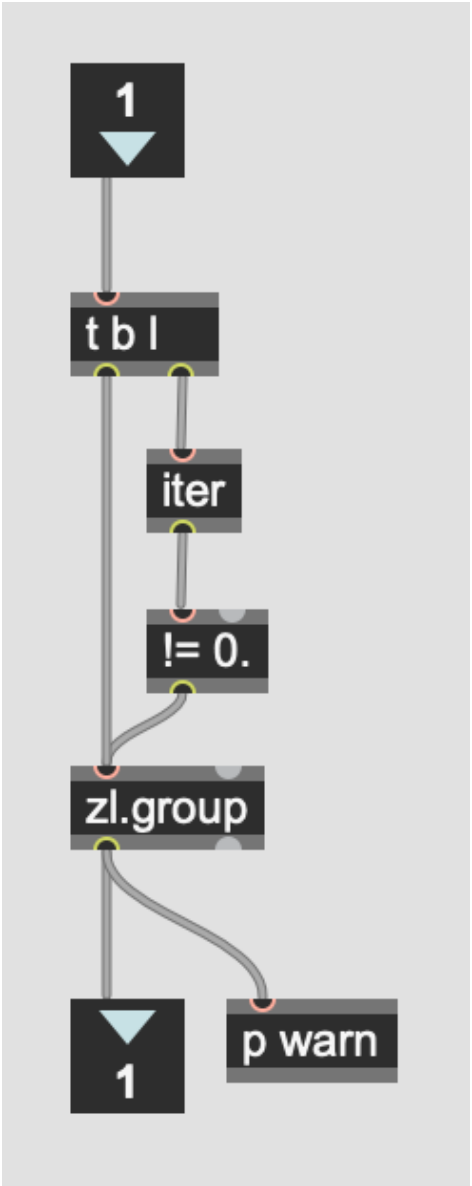

Figure 21 - Contents of *get_target_states* patcher

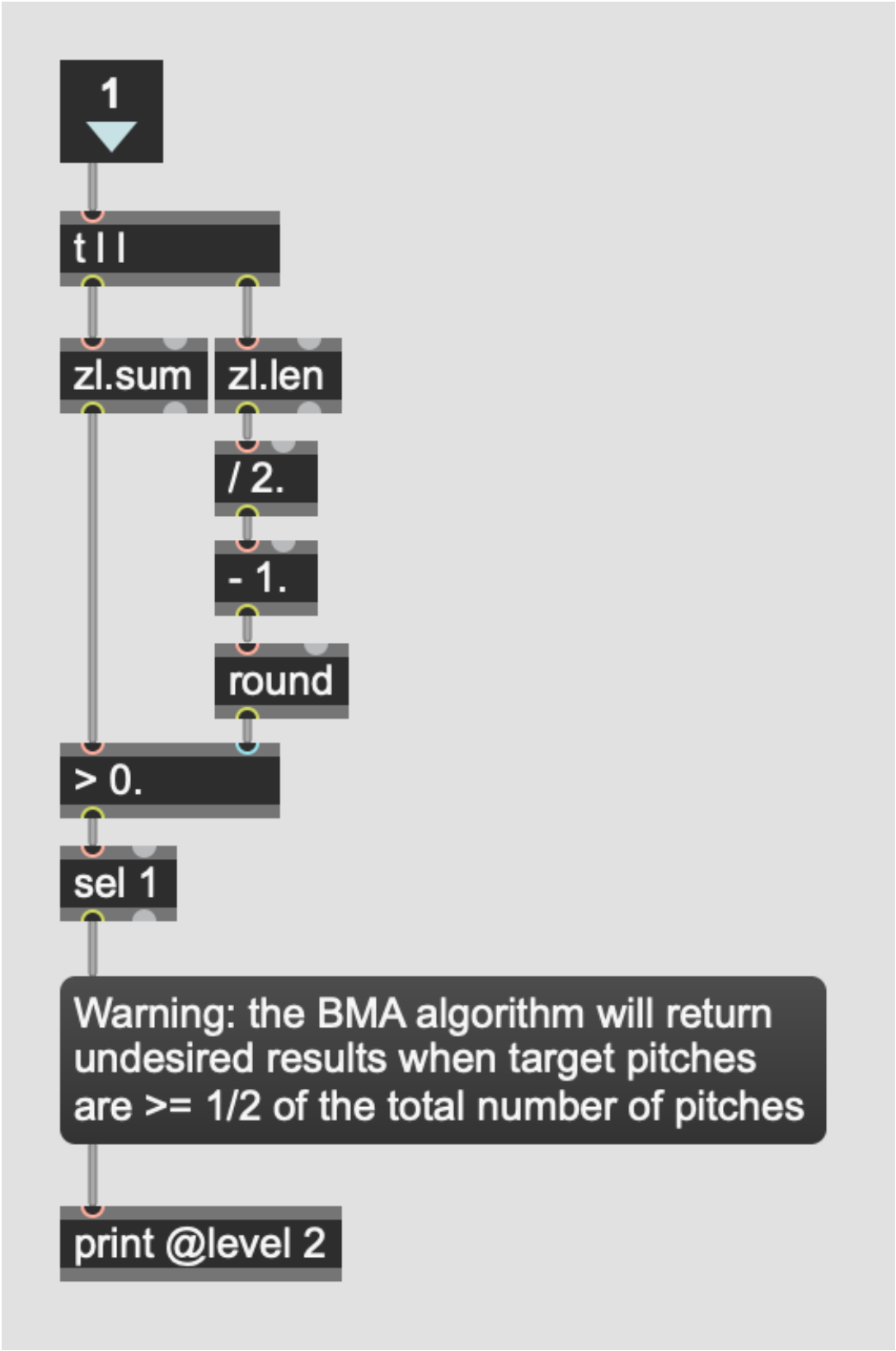

Figure 22 - Contents of *warn* patcher

This new list of zeros and ones, that represent the target pitches (or states) that can follow the selected initial pitch, according to the transition table provided, is then passed onto the *build_circuit* patcher that will, as the name suggests, programmatically build the quantum circuit. This is perhaps the most important patcher in this whole implementation, and it is where I am mostly using the tools provided by *The QAC Toolkit* (see figure 23). The list of target states will first be translated by the *states2qubits* patcher into an integer number that is equal to the number of qubits required to represent the number of states in our list. In this case, we have 12 pitches, so we will need 4 qubits to be able to represent them. Looking at this patch as being organized in 7 horizontal layers, we can see that this calculated number of required qubits is temporarily stored in mostly message boxes on the 4th horizontal layer[37] that will be triggered next.

[37] Here I am counting these horizontal layers from the top. As a related note, in Max the order of operations happens from top to bottom, from right to left.

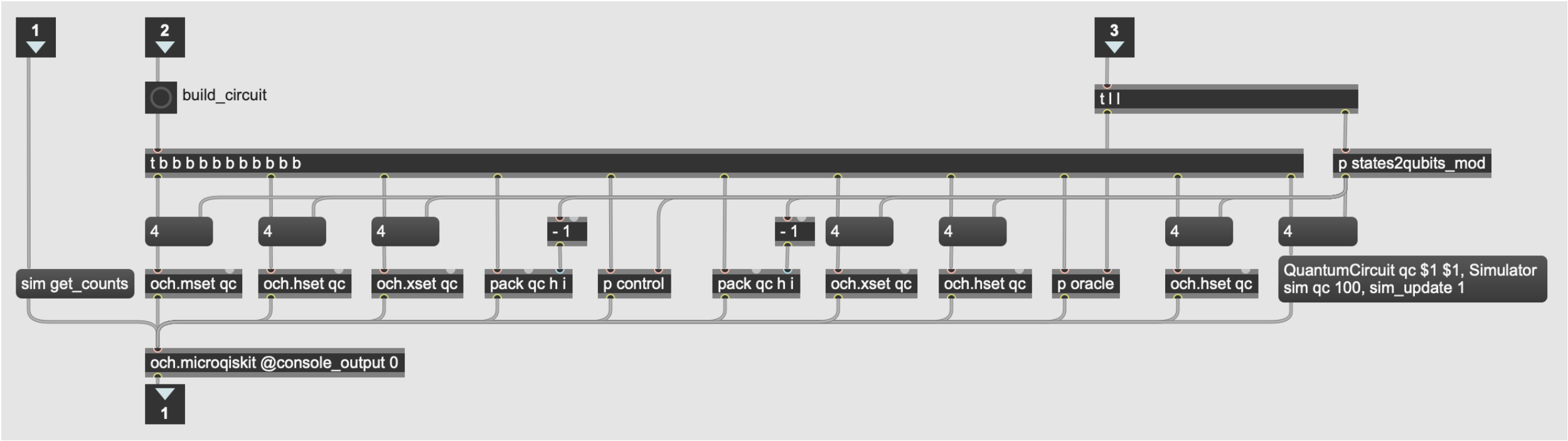

Figure 23 - Contents of *build_circuit* patcher

The *bang* triggering the actual automatic circuit building is coming from the trigger object in the main patch that will fire a *bang* as the last action after receiving the initial note from the *umenu*. This *bang* is then split into 11 different *bangs* that are fired consecutively, by the large trigger object inside the *build_circuit* patcher on the 3rd horizontal layer. In turn, these bangs are directly responsible for generating all the messages to control the **och.microqiskit** object using the different objects in the 5th horizontal layer, in the right to left order. The first message is setting up (or recreating) a *QuatumCircuit* named *qc* with the same number of qubits and classical bits (determined by number received on its inlet), as well as creating a *Simulator* named *sim* with 100 shots that will hold the circuit named *qc* and auto-update its memory when *qc* changes.

Following the definition of BMA, a series of hadamard gates is created covering the entire range of qubits in the circuit, and we use **och.hset** to make this process faster. Giving an argument to this object that is the name of the *QuantumCircuit* in question, in this case *qc*, and sending it an integer number, representing the number of qubits of the circuit, this object will automatically generate a series of messages that will fit our need - i.e. “qc h 0,” “qc h 1,” “qc h 2,” and “qc h 3.” The **och.xset** and **och.mset** objects, that are found later in this patch, work exactly the same way.

The oracle portion of BMA is built next (see figure 24). This is entirely generated using only the transition table row transformed by the *get_target_states* patcher before. After creating an identity matrix of the required size using **och.identitymatrix** object, the patch starts iterating through its multiple rows at the same time that it is reading each of the values from the transition table row above. When the value for the transition table row evaluates to 1, then the corresponding row from the identity matrix is opened and have its value multiplied by -1. This process goes on until the transition table row reading is complete, and all the resulting iterations and modifications are grouped together into a new matrix.

Again, given that the total number of pitches in the original transition table might be less than the number of possible states that the qubits can represent, the patch will still need to stitch together a *tail* of an identity matrix long enough to complete the full matrix - see the trio of *zl* objects at the bottom right section of the patch.[38] All throughout this patch, matrices take the form of a list of symbols, where each symbol is itself a list of values. In order to request

[38] In this case, the 12 pitches require 4 qubits, and the 4 qubits can represent 16 states. We need to complete the 12 x 12 matrix in order to become a full 16 x 16 matrix that can be applied to the quantum circuit covering the full range of qubits.

och.microqiskit to create a unitary gate from a matrix, though, the standard message format must be maintained: *QuantumCircuit* name, quantum gate type, and list of arguments. This is why the final complete matrix is transformed into a long list of single values before it is passed to the last message at the bottom of the patch.

The rest of the circuit building will either repeat or use very similar strategies for programmatically generating the required messages. Only the *control* subpatcher will offer 3 different options for its generation (see figure 25). At some point in BMA there is a controlled-not gate, or multi-controlled-not gate, being applied to the quantum circuit. As it stands, my implementation is prepared to select the appropriate gate when the number of qubits of the quantum circuit is between 2 and 4. It should be noted that a 1 qubit version of BMA cannot exist: if we have a transition table between 2 pitches, it may seem that we only need 1 qubit to represent the required states but, as noted above, by design BMA will not retrieve the desired results if targeting 1 state out of 2 possible states. Again, the proportion between target states and total number of possible states needs to be less than ½.

With the quantum circuit defined for the initial pitch selected on the main patch, all that is left is running the actual program and retrieving some results. The *start* toggle on the main patch initiates a loop that will repeat a number of times determined by the *num_loops* number box. The *counter* and *select* objects are keeping track of the repetitions, and as long as the maximum number of repetitions isn't reached, it will keep sending *bangs* to the first inlet of *build_circuit*. Because at this point the *QuantumCircuit* and *Simulator* is already completely defined, all that is left to do is request **och.microqiskit** object to simulate the circuit and retrieve the counts - this is done by the "sim get_counts" message.

The counts are then passed to the *calc_next_note* patcher that will unpack the list of results, sort them, and select the bitstring for the state with the highest number of counts - see figure 26. This bitstring is then converted into a decimal number with the **och.binstr2dec** object, and this number is used to retrieve both the MIDI note number as well as pitch name from the initial list of labels that was stored in the *next_notes* value object at the start. The MIDI note number is sent out via the third outlet to a MIDI instrument synthesizer, and the pitch name is sent via the second outlet back into the trigger object, on the main patch, that is responsible for programmatically generating the quantum circuit, as described in detail in this section.[39] After that, a final *bang* is sent out via the first inlet and will trigger, 150ms later, the next iteration of the loop.

This patch is available in [49] and I invite the reader to explore it, including changing the initial rules and labels that are defined at the start. As was explained in this section, rather than hardcoding the exact example provided by Basak and Miranda, this patch took several steps in consideration in order to make it somewhat flexible and able to adapt to different transition tables which might have different sizes, and thus also produce different quantum circuits. In the next section, I will explore a different approach to QAC, still making use of *The QAC Toolkit*, in the context of the creative work *Disklavier Prelude #3*.

[39] Note that a different pitch name entering this trigger object will result in the extraction of a different row from the transition table and, consequently, in the generation of a different quantum circuit.

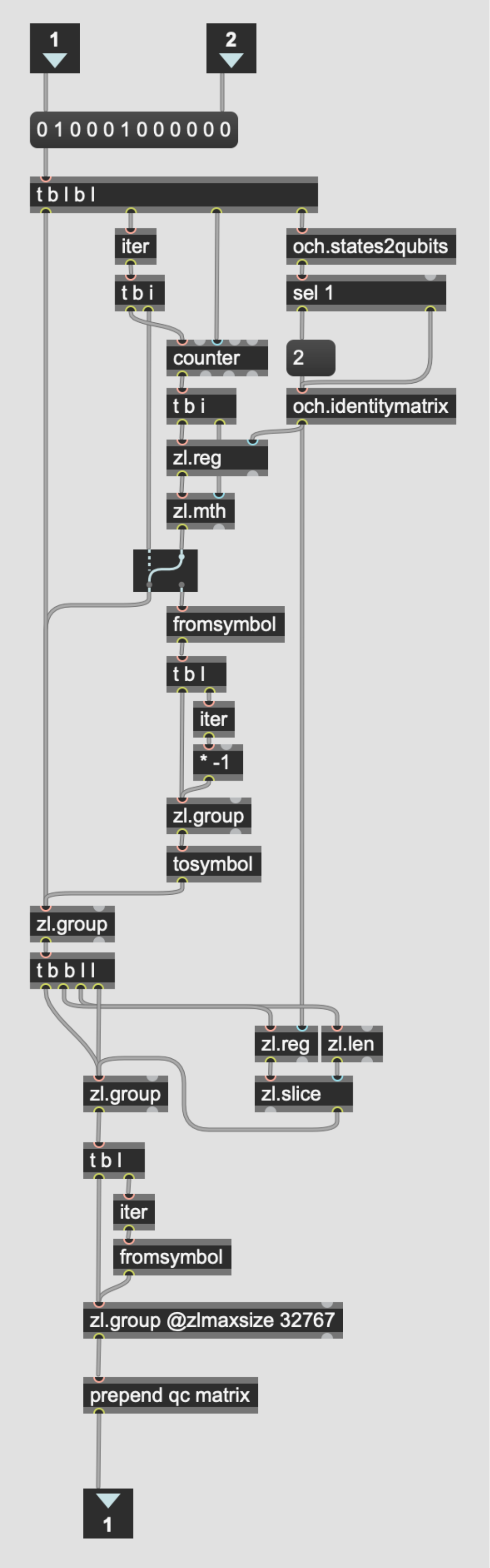

Figure 24 - Contents of *oracle* patcher

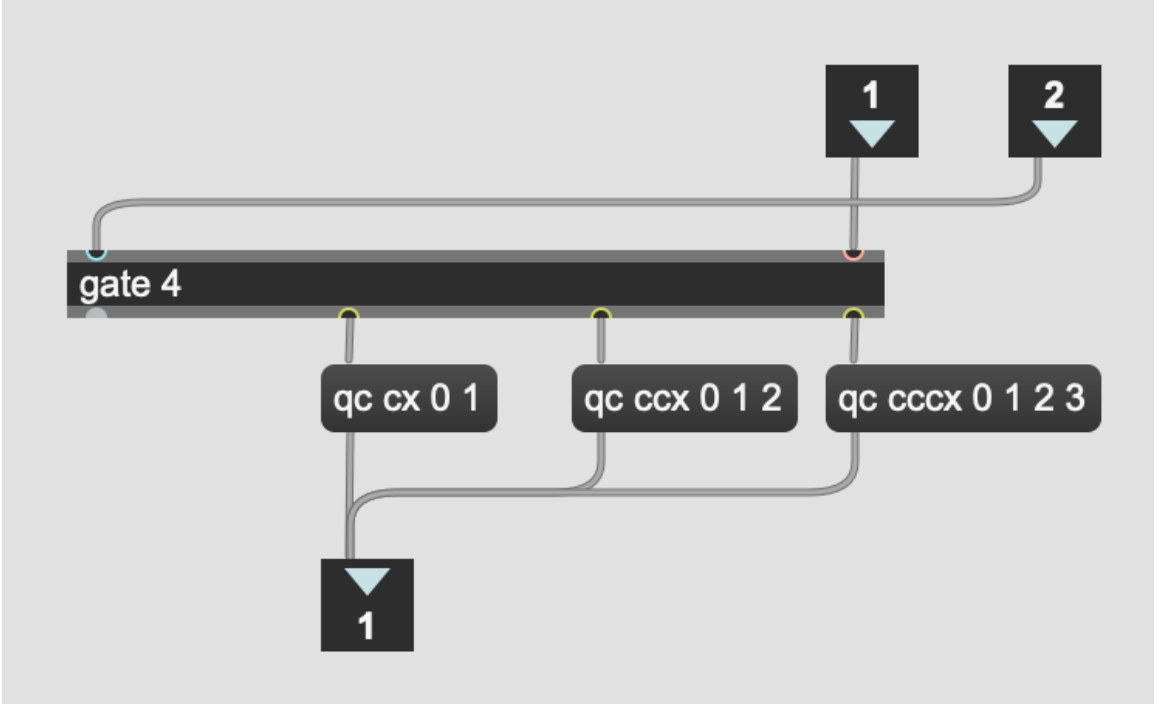

Figure 25 - Contents of “control” patcher

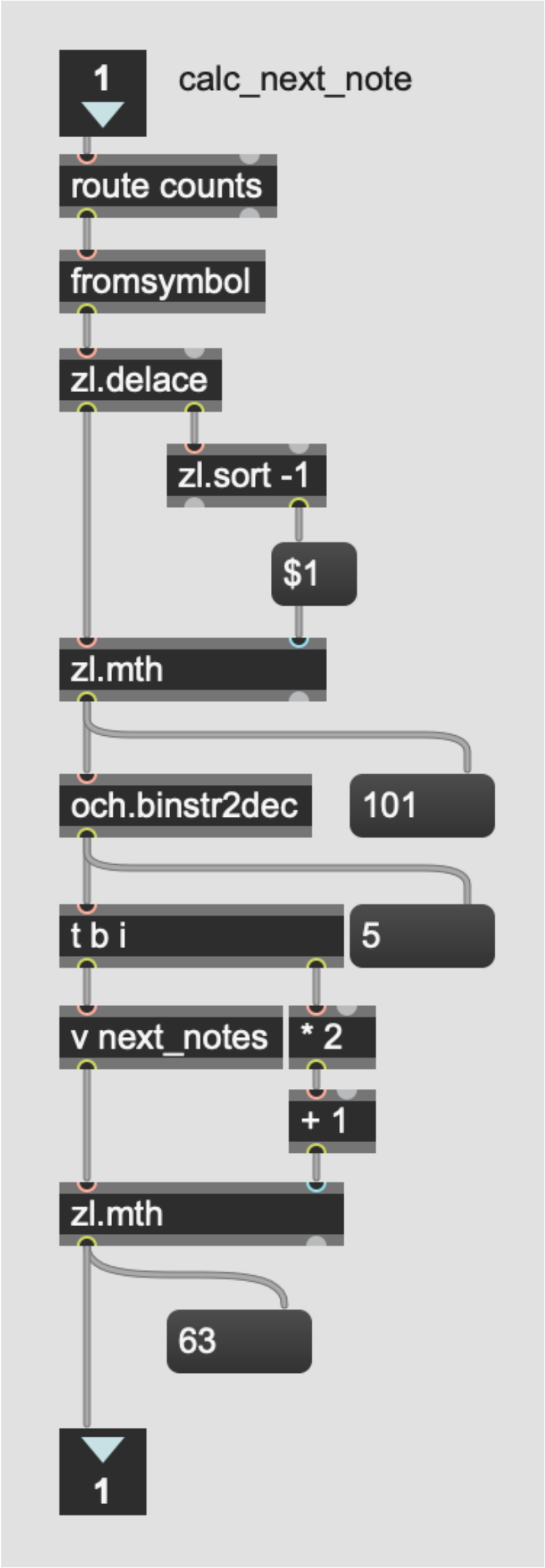

Figure 26 - Contents of “calc_next_note”

## QAC in *Disklavier Prelude #3*

In a nutshell, *The QAC Toolkit* is what I wish would exist back in the early 2010's when I first became aware of quantum computing, and especially when I was struggling to understand and integrate QCL. But this process doesn't end here. Now, given that there is such a toolkit, I can create compositional devices that make use of it, but don't necessarily require me to interact with them at the level of programming patches with max objects and messages. One of the main advantages of using Max, as mentioned before, is the ability to quickly deploy plugin-like Max for Live devices that act as ready-to-use plugins in the DAW environment Ableton Live. When using the DAW, I am mostly working in a *creative mode* and less of a *programmer mode*.[40] The ability to quickly put together a set where I can connect virtual and acoustic instruments, and make use of both traditional techniques and new QC-based ones, is truly invaluable.

One of the first devices of this kind, to include and make use of *The QAC Toolkit* objects under the hood, is **och.qc-superposition**.[41] This Max for Live device presents the user with a simple interface that abstracts away the process of writing the actual quantum circuit (see figure 27). On the bottom portion, there is a visualization of the counts retrieved from running a quantum circuit, as an histogram. And on the top portion, there is a number box to determine the number of qubits (and classical bits) to allocate, a *Simulate* button to trigger the simulation, a text field that declares the top ket (as in the wining computational basis state), and a mappable *result* knob that reflects the state with highest number of counts too.

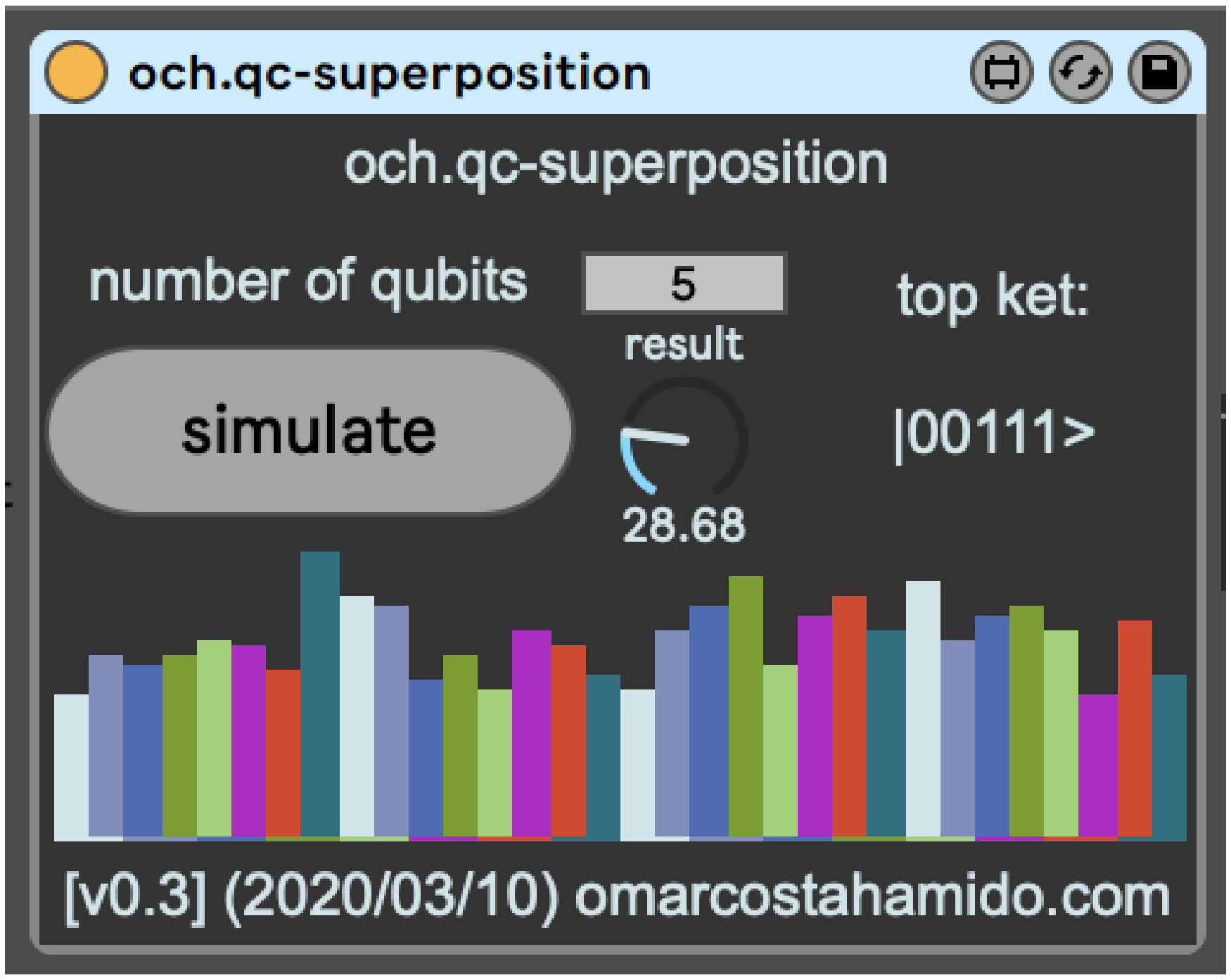

Figure 27 - The och.qc-superposition Max for Live device, using *The QAC Toolkit*.. From [27]

[40] By the same token, it also makes these tools more accessible to non-programmers.
[41] This device is included in *The QAC Toolkit* Max package [1].

Internally, **och.qc-superposition** creates a simple circuit with a hadamard gate and a measurement gate applied to each qubit, effectively putting all qubits in a state of equal superposition. By changing the number of qubits, the total number of possible states increases. In practical terms, this change is noticeable as a change in resolution of the *result* knob, that will be able to output finer variations as the number of qubits increase: its range is divided by $2^{(\text{number of qubits})}$ parts. This device represents a fundamental departure from the constraints encountered previously with **och.qc-circ-2q**. At the same time, it also represents QC more faithfully, as a probabilistic means of computation. Instead of simply retrieving a statevector, we are actually simulating a circuit, and that drives the change in this device.

One of the music compositions that make use of this device is *Disklavier Prelude #3*.[42] This is a work for Disklavier, dancer, and lights that explores the concept of virtual particles, QC, man-machine interaction, and social media (see figure 28). All the way through the piece, the Disklavier plays maj7 chords that are transposed up and down based on the result of the **och.qc-superposition** device. At some point during the piece, the lampshade that the dancer is wearing falls off, and the dancer becomes aware of her surroundings and the controlling light. At this point the Disklavier adds a melodic line to the arpeggiated chords that creates a dialogue with the dancer. The melodic line, that appears to be trying to communicate with the dancer, is composed of streams of notes from a pentatonic scale being transposed together with the chords, and being driven by another **och.qc-superposition** device.

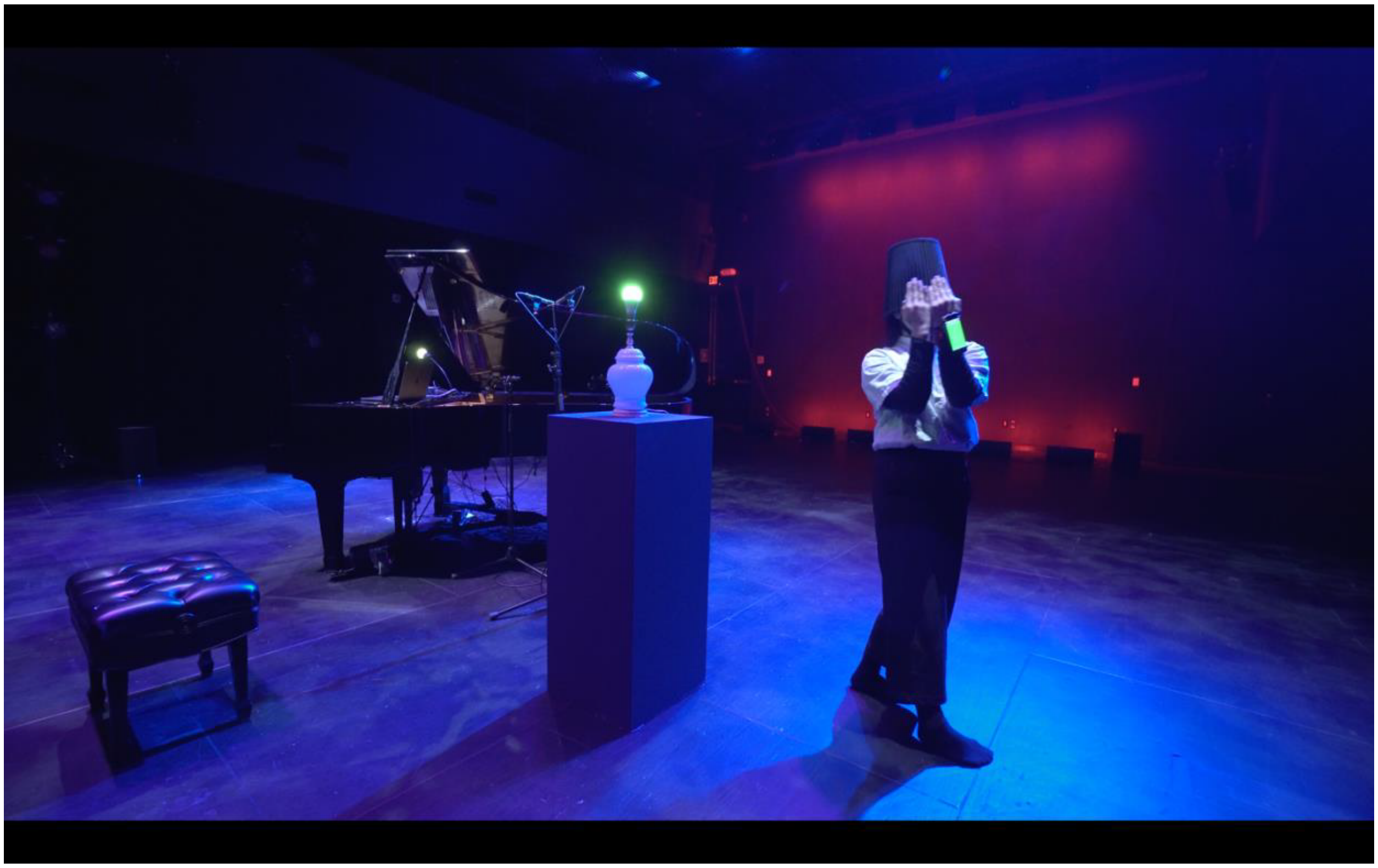

Figure 28 - Still frame of *Disklavier Prelude #3*. From [62]

[42] Part of the work *4 Disklavier Preludes*, presented in the film-recital *The Gedanken Room* (2021) [62]

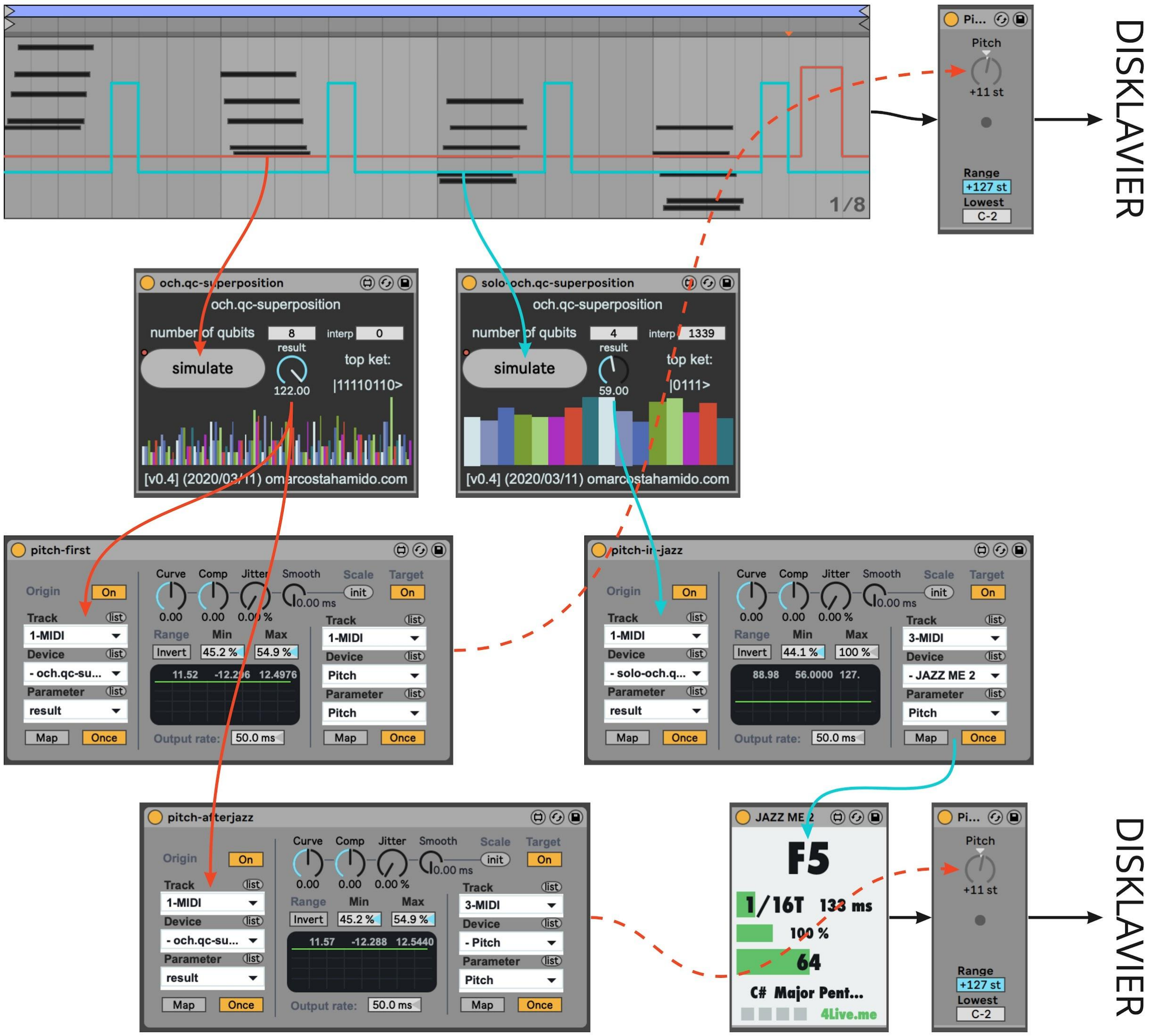

Figure 29 - One automation line triggers och.qc-superposition to control transposition, the other automation line triggers another och.qc-superposition to control a stream of notes on *Jazz me.* From [27]

Figure 29 illustrates in more detail the compositional process involving the use of this device. There is a MIDI clip with arpeggiated maj7 chords driving the Disklavier that includes 2 automation lanes for triggering each of the **och.qc-superposition** device instances. The first automation lane only triggers a new simulation at the end of the clip, on a device instance configured with a high qubit count in order to achieve a finer resolution while controlling the transposition. The devices were modified here to include an interpolation time, to smooth out the transitions. On the first instance interpolation is set to zero because those changes should take effect immediately. The actual transposition is achieved simply by mapping the *result* knob to the *pitch* knob, on a *Pitch* device, that transposes the MIDI information being sent to the Disklavier.

The second automation lane, that is enabled later in the piece when the lampshade falls off, is responsible for controlling the melodic stream of notes. This automation triggers, every bar, a second **och.qc-superposition** device configured with a lower qubit count and a much higher interpolation time. The *result* knob then slowly transitions between results every time it is triggered, and this movement is mapped to an instance of Yehezkel Raz's *Jazz me* device that generates a stream of notes in a major pentatonic mode. These notes are guaranteed to stay in tune with the arpeggiated chords because they are also being transposed by the first automation lane chain.

This complex network of interdependencies, including in the synchronized lights, allowed to express a very controlled scenario of a seemingly unknown outcome. Since the QC process is relying on a simple circuit with qubits in equal superposition, any possible value is equally likely. The predetermined automation lanes, resembling the control pulse automations generated from quantum circuits when passed to quantum computers,[43] depict these tensions between control and uncertainty. And, in turn, the repeating rhythmic motif was also very important in order to be able to articulate well with the dancer, by allowing her to express the constraints, as well as break free from them, and freely explore this creative environment.

## Closing remarks

In this chapter I presented QAC (Quantum-computing Aided Composition) as a proposal to integrate QC in creative practice. This is a contribution emerging from the *Adventures in Quantumland* research [27].[44] In particular, QAC proposes to draw a framework that places the creative artist at the center of this new landscape, revealed in the exploration of the paradigm introduced by QC.

It is important to understand the context and the relations between CM and QC, as well as to know what QC is and what can be expected from it. The fact that there is no real (perceptible) *quantum sound* may lead to the question of whether an audience member will ever be able to determine if some sound, or music, was produced using QC - perhaps only the composer or performer using the system will be able to know, really.

Previous attempts for an integration of QC with music practice tend to favor code-based languages and require a skillset that only the more programmer-inclined creative artists possess. Being informed by realtime creative practices, QAC places the emphasis on the creative process, and interaction with performers, rather than in exclusively computer-driven artwork.

*The QAC Toolkit* [1], as well as some early attempts before it, emerged within this context to empower creative artists to use quantum gates as logical operations in their practice. As a modern attempt to put QAC into practice, it provides new tools integrated into the Max visual programming environment with a performance that, even though it might not be relevant to the computer scientist, may enable new prospects for realtime creative practice with QC.

---

[43] The quantum circuit, as a set of instructions, is actually converted in a series of pulses to control the real quantum hardware. Pulse scheduling is involved in the actual implementation of each quantum gate, and it is one of the avenues where researchers attempt to optimize QC and mitigate noise.

[44] Where I also proposed QAD (Quantum-computing Aided Design).

Still, the **och.microqiskit** object includes methods to retrieve the circuit definition in Qasm, using *get_qasm*, and Qiskit Python code, using *get_qiskit*, offering the possibility to explore and reuse the current circuits in other environments.[45] With this, and during the research process involved in the creation of these tools, there were some contributions to the QC community. For example, the version of MicroQiskit that I've worked on is the only one to include multi-qubit gates other than crx and cx.

Furthermore, the examples described in this chapter demonstrated how these tools can be used when creating new patches, implementing algorithms, building plugin-like devices, and composing a new musical work. The ability to abstract away the underlying QC process inside the self-contained devices makes it accessible to non-programmers and a wider audience of creative practitioners.

It is my intent that the tools, concepts, and perspectives shared here will be inspiring to both the creative and scientific community, promoting dialogues about their respective practices and possible intersections.

[45] These can be easily copied from a popup text box that appears when the method is called in a message that includes the keyword textbox (e.g. "sim get_qiskit textbox").